\documentclass[pdflatex,sn-mathphys-num]{sn-jnl}% Math and Physical Sciences Numbered Reference Style 

\usepackage[mathlines]{lineno} % numbers lines in math too
\usepackage{graphicx}%
\usepackage{multirow}%
\usepackage{mathtools}
\usepackage{amsmath,amssymb,amsfonts}%
\usepackage{amsthm}%
\usepackage{mathrsfs}%
\usepackage[title]{appendix}%
\usepackage{xcolor}%
\usepackage{textcomp}%
\usepackage{manyfoot}%
\usepackage{booktabs}%
\usepackage{algorithm}%
\usepackage{algorithmicx}%
\usepackage{algpseudocode}%
\usepackage{listings}%
\usepackage[T1]{fontenc}
\usepackage{xcolor}
\usepackage{pdfpages}
\usepackage{soul}
\soulregister\ref7
\soulregister\cite7
\usepackage[normalem]{ulem}
\theoremstyle{thmstyleone}%
\theoremstyle{thmstyletwo}%

\theoremstyle{thmstylethree}%

\begin{document}
%\linenumbers                 % turn on numbering
%\title[Article Title]{Near unity polarization-independent deterministic beam localization in a PL}

\title[Article Title]{Geometric scaling limits of phase-only control in multimode coherent systems}

\author[*,1]{Harikumar K Chandrasekharan}

\affil[1]{Scottish Universities Physics Alliance, Institute of Photonics and Quantum Sciences, School of Engineering
and Physical Sciences, Heriot-Watt University, David Brewster Building, Edinburgh EH14 4AS, Scotland, UK}

\affil[*]{hk47@hw.ac.uk}

\abstract{Control of coherent waves is often restricted to phase-only actuation in multimode systems, yet the resulting physical limits remain poorly understood. Here, we show that restricting control to relative phases confines dynamics to a compact manifold whose geometry produces isolated stationary interference basins with robustness governed by local curvature. Imperfections act as smooth perturbations that soften basin structure without eliminating stationary states. This geometry imposes a universal scaling constraint: although the number of stationary states increases with system dimensionality, achievable localization contrast degrades through leakage into uncontrolled degrees of freedom. Experimentally, we demonstrate this in a telecom-wavelength multimode photonic lantern, where coarse phase-only scans directly map stable interference basins, reveal efficiency–stability trade-offs, and identify robust operating regimes without transmission-matrix reconstruction, adaptive optimization, or system inversion. The framework establishes a practical calibration-free approach for constrained multimode coherent control and applies broadly to optical, microwave, acoustic, and finite-dimensional quantum systems.}

%\keywords{keyword1, Keyword2, Keyword3, Keyword4}



\maketitle
\newpage
\section{Introduction}\label{sec1}
In coherent wave systems, interference among coupled degrees of freedom determines whether energy remains distributed or concentrates into spatially localized configurations \cite{2000stop.book}. This principle underlies diverse platforms, including multimode optical fibers, integrated photonics, acoustic cavities, and microwave resonators \cite{Gloge:71,articleDJ,Miller:13,PhysRevLett.75.4206,PhysRevLett.92.193904}. Substantial progress in controlling such systems has been achieved through wavefront shaping, phase conjugation, and transmission-matrix approaches \cite{articlemosk,PhysRevLett.104.100601,1.5136334,Vellekoop:15,YU2022100292,Vellekoop_2008}, enabling refocusing through strongly mixed media \cite{Papadopoulos:12,9420070} and compensation of modal dispersion and disorder \cite{MORI2013132}. These approaches, however, typically rely on high-dimensional control, detailed system knowledge, or iterative optimization, leaving open what limits arise under intrinsically constrained control.

Phase-only actuation over a finite set of modes is a common constraint, yet its structure remains poorly understood. Established approaches—including transmission eigenchannels~\cite{articlemosk,RevModPhys.89.015005}, principal modes~\cite{PhysRevX.7.041053}, and scattering-invariant states~\cite{PaiNatPhoton2021}—identify privileged configurations in complex systems but generally assume unconstrained control. By contrast, restricting actuation to relative phases confines dynamics to a compact $(N-1)$-dimensional manifold, where phase-only control resides on a torus whose topology and curvature govern the multiplicity and stability of accessible interference states~\cite{Miller:13}. Unlike control theory and control landscapes defined over unconstrained complex amplitudes, this compactness enforces discrete stationary configurations and fundamentally alters the structure and scaling of accessible states.

This geometric confinement introduces a fundamental tension. While low-dimensional modal recombination can exhibit persistent localized states, their fidelity typically degrades with increasing dimensionality or imperfection \cite{PhysRevX.7.041053,10.3389/fphy.2021.713085}. Existing bounds on phase-only focusing quantify asymptotic enhancement at a target \cite{64xd-3cbs,GHANGAS2025131876}, but assume optimization over the full complex coefficient space and do not describe how stationary configurations are organized on a compact phase manifold with fixed amplitudes, nor how their robustness scales with system size. A central question therefore remains: \emph{what universal constraints govern localization and robustness under phase-only control?}

Motivated by our coherent mode recombination experiments \cite{chandrasekharan2025}, where phase scans reveal discrete and reproducible localization states, we identify a geometric origin of these limitations. Confinement to a compact phase manifold imposes intrinsic constraints: linear coherent systems generically support isolated stationary interference configurations with finite basins of attraction, whose robustness is set by local curvature. Imperfections---including loss, residual mixing, and leakage---act as smooth deformations of an effective Gram operator \cite{Gloge:71,RevModPhys.89.015005}, softening basin curvature and limiting achievable contrast.

The framework is platform-independent and applies to coherent systems described as finite modal superpositions with controllable relative phases \cite{Miller:13}. As a concrete realization, we consider the mode-selective photonic lantern \cite{Birks:15,Leon-Saval:14,Sai:17,2012MNRAS,Lin:25,Becerra-Deana:25}, which provides direct experimental access to compact phase-space geometry and scaling-induced basin softening. Increasing modal dimensionality does not eliminate stationary interference states; instead, unavoidable asymmetry reduces basin curvature, imposing a geometric ceiling on recombination performance.

Crucially, robustness is governed by geometric quantities intrinsic to the compact control manifold and directly measurable without transmission-matrix reconstruction or iterative optimization. Basin curvature follows from quadratic degradation of a localization metric under small phase perturbations, requiring only phase modulation and scalar readout. Unlike conventional wavefront-shaping approaches, this enables direct identification of stable, high-performance states with minimal measurement overhead. Experimentally, coarse voltage-driven phase scans—where voltage acts as a proxy for relative phase—provide a simple route to identifying efficient and robust operating regimes without full system calibration. This establishes a practical framework for operating multimode systems under constrained control. The same compact-manifold structure governs phase-constrained coherent control across optical \cite{1.5136334}, microwave \cite{PhysRevLett.92.193904}, acoustic \cite{Fink1997}, and finite-dimensional quantum systems \cite{Chakrabarti01102007}.
   
\section{Universal phase-space framework for constrained multimode control}
\label{sec:model}

We establish a minimal geometric framework for phase-only control in coherent multimode systems. When only relative phases are tunable, physically distinct configurations lie on a compact manifold whose geometry—rather than microscopic detail—governs interference and stability. This compactness generically produces discrete stationary localization states, whose robustness is set by local curvature, while imperfections impose a dimensional scaling limit.

\begin{figure*}[b]
\centering
{\includegraphics[width=\linewidth]{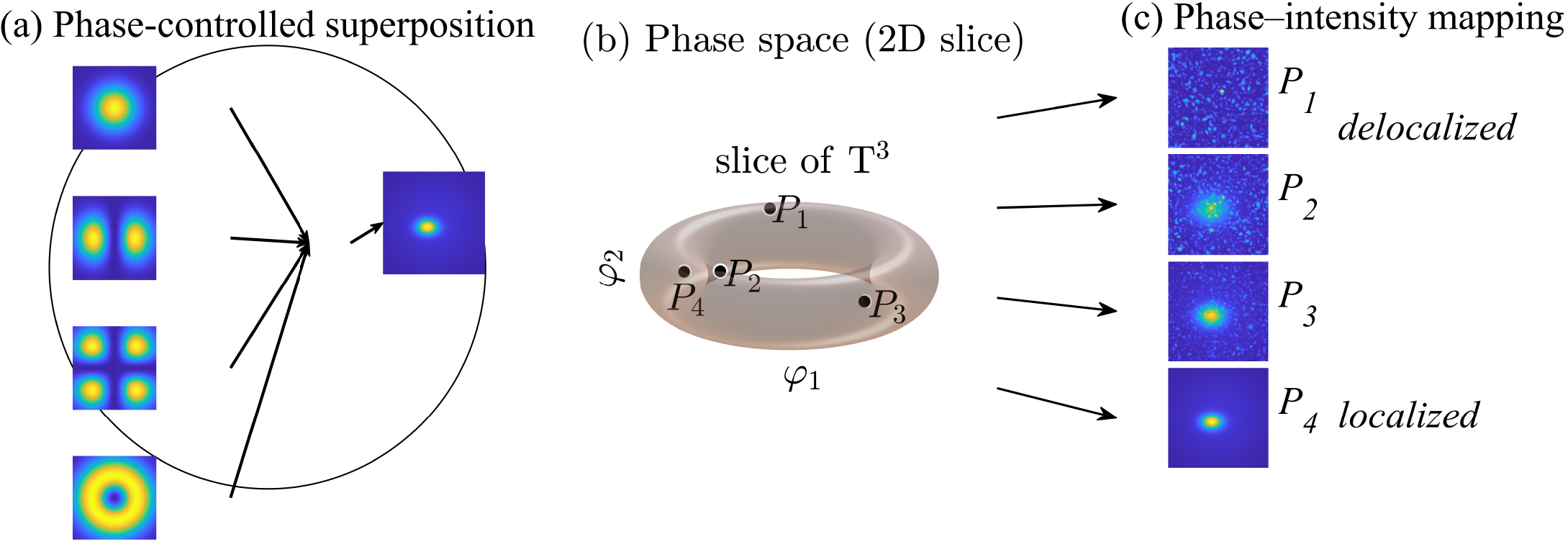}}
\caption{\textbf{Phase-controlled multimode interference and phase-space representation.} \textbf{(a)} A coherent field is synthesized as a superposition of guided modes with fixed amplitudes $c_m$ and independently tunable phases $\{\phi_m\}$, producing an output field $E(\mathbf{r})$ through multimode interference. \textbf{(b)} Removing the global phase redundancy yields an $(N\!-\!1)$-torus $\mathbb{T}^{N-1}$ of reduced (relative) phases $\boldsymbol{\varphi}$ (for $N=4$ this is $\mathbb{T}^{3}$). The reduced phases are defined as $\varphi_{m-1} \equiv \phi_m - \phi_1$ for $m=2,\dots,N$, corresponding to a choice of absolute reference phase. For visualization, we show a two-dimensional slice of this space parameterized by $(\varphi_1,\varphi_2)$—i.e.\ $\varphi_1=\phi_2-\phi_1$ and $\varphi_2=\phi_3-\phi_1$—with $\varphi_3$ held fixed ($\varphi_3=\phi_4-\phi_1$), matching the phase coordinates in panel, and indicate four representative phase points $P_1$--$P_4$. \textbf{(c)} The spatial output intensity patterns corresponding to the phase points $P_1$--$P_4$ in panel (b), illustrating the evolution from delocalized (speckle-like) to localized output states under phase-only control.}
\label{fig1}
\end{figure*}

\medskip
\noindent\textbf{Phase-only control and reduced phase-space.}
A general coherent multimode output field can be written as a finite superposition of spatial modes~\cite{Gloge:71}, \begin{equation} E(\mathbf{r};\boldsymbol{\phi}) = \sum_{m=1}^{N} c_m\,u_m(\mathbf{r})\,e^{i\phi_m}, \label{eq:modal_expansion} \end{equation} where $u_m(\mathbf{r})$ denote mode functions, $c_m$ are fixed complex amplitudes encoding excitation weights and system transfer coefficients, and $\phi_m$ are externally controlled phases. This representation applies broadly to coherent wave platforms, including multimode waveguides and resonant cavities.

As illustrated in Fig.~\ref{fig1}a, spatial control arises from tuning the relative phases of fixed modal contributions. Removing the global phase reduces the independent control parameters to an $(N\!-\!1)$-dimensional torus $\mathbb{T}^{N-1}$ (Fig.~\ref{fig1}b). Each point on this compact manifold defines a distinct coherent superposition and corresponding output intensity pattern (Fig.~\ref{fig1}c). Importantly, Eq.~(\ref{eq:modal_expansion}) does not assume orthogonality, equal normalization, or lossless propagation: static mixing, loss, or amplitude imbalance are absorbed into the coefficients $c_m$ and effective mode profiles, provided coherence is preserved (see Supplementary Note~1).

\begin{figure*}[t]
\centering
{\includegraphics[width=0.9\linewidth]{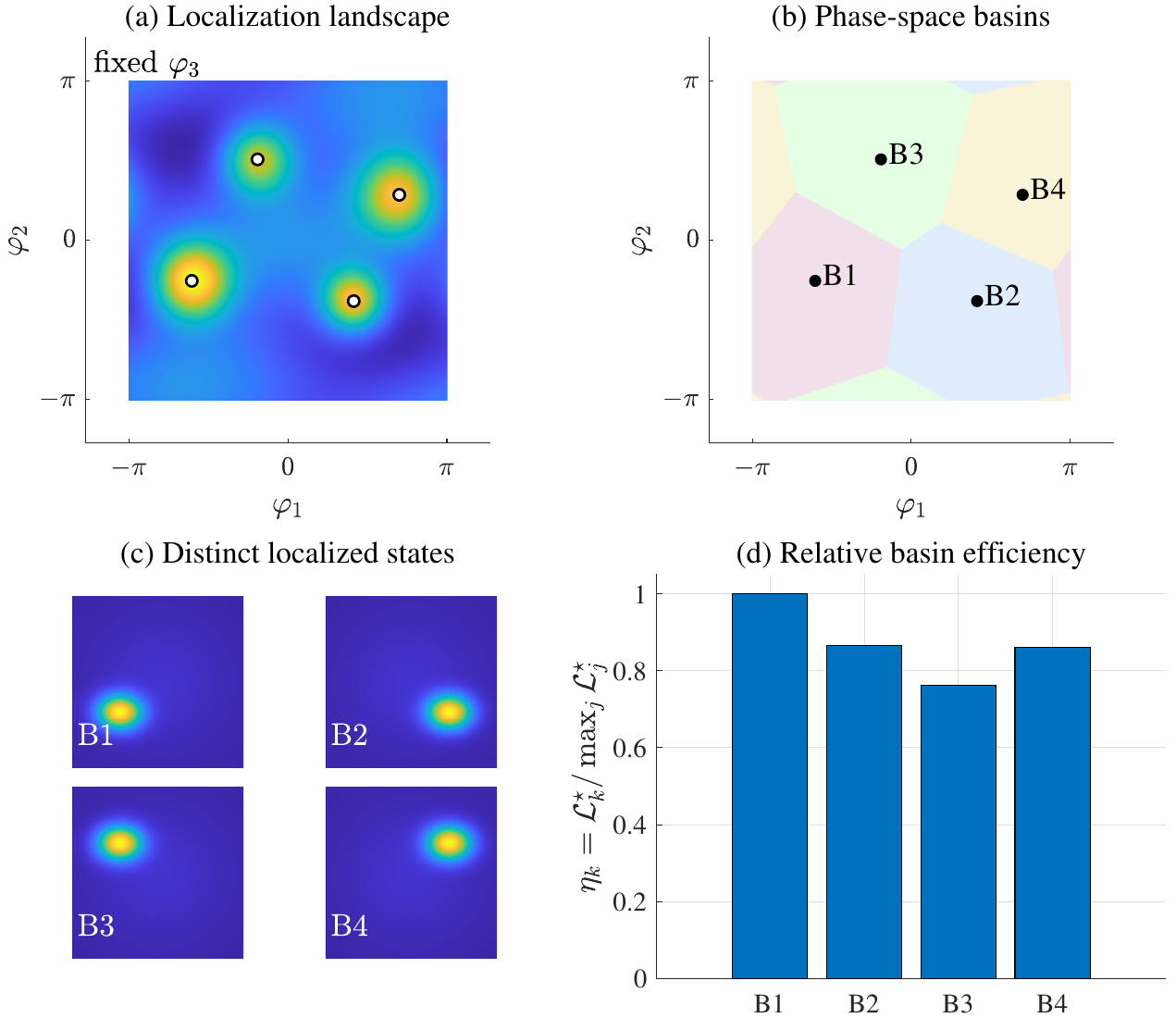}}
\caption{\textbf{Phase-induced localization landscapes and basin structure in a four-mode system.} \textbf{(a)} Localization functional $\mathcal{L}(\boldsymbol{\varphi})$ evaluated over a two-dimensional slice of the reduced phase-space $\mathbb{T}^{3}$ (with $\varphi_3$ held fixed), shown as a smooth phase-space landscape with multiple local maxima. These extrema define distinct basins of optimality on the phase-space landscape, indicated by the labeled points $B_1$--$B_4$. \textbf{(b)} Basin decomposition of the same phase-space slice, illustrating how the localization landscape partitions the control manifold into regions associated with different stable localized states. \textbf{(c)} Representative spatial output intensity patterns corresponding to the basins $B_1$--$B_4$. The correspondence between panels is indicated by basin labels rather than by spatial position: panel (a) encodes phase-space geometry, while panel (c) displays the identities of the localized states associated with each basin. \textbf{(d)} Relative basin efficiency, showing the maximum localized energy achieved within each basin, normalized to the highest localized energy across all sampled basins.}
\label{fig2}
\end{figure*}

\medskip
\noindent\textbf{Localization functionals and phase-space landscapes.}
To characterize spatial localization, we introduce localization functionals that assign a scalar measure of concentration to a given output field. Evaluating the localization functional $\mathcal{L}$ over the compact phase space $\mathbb{T}^{N-1}$ defines a smooth landscape $\mathcal{L}(\boldsymbol{\varphi})$, whose compactness enforces isolated basins absent in unconstrained control. As shown in Fig.~\ref{fig2}a, this landscape exhibits multiple local maxima that partition the reduced phase space into distinct basins of attraction (Fig.~\ref{fig2}b), each corresponding to a localized output pattern (Fig.~\ref{fig2}c); Fig.~\ref{fig2}d compares their relative peak values. The landscape dimensionality is determined solely by the number of controllable phases, even though the physical field may inhabit a much higher-dimensional space. Formal definitions of localization functionals, basins, and robustness neighborhoods are given in Supplementary Note~2.

\begin{figure*}[t]
\centering
{\includegraphics[width=\linewidth]{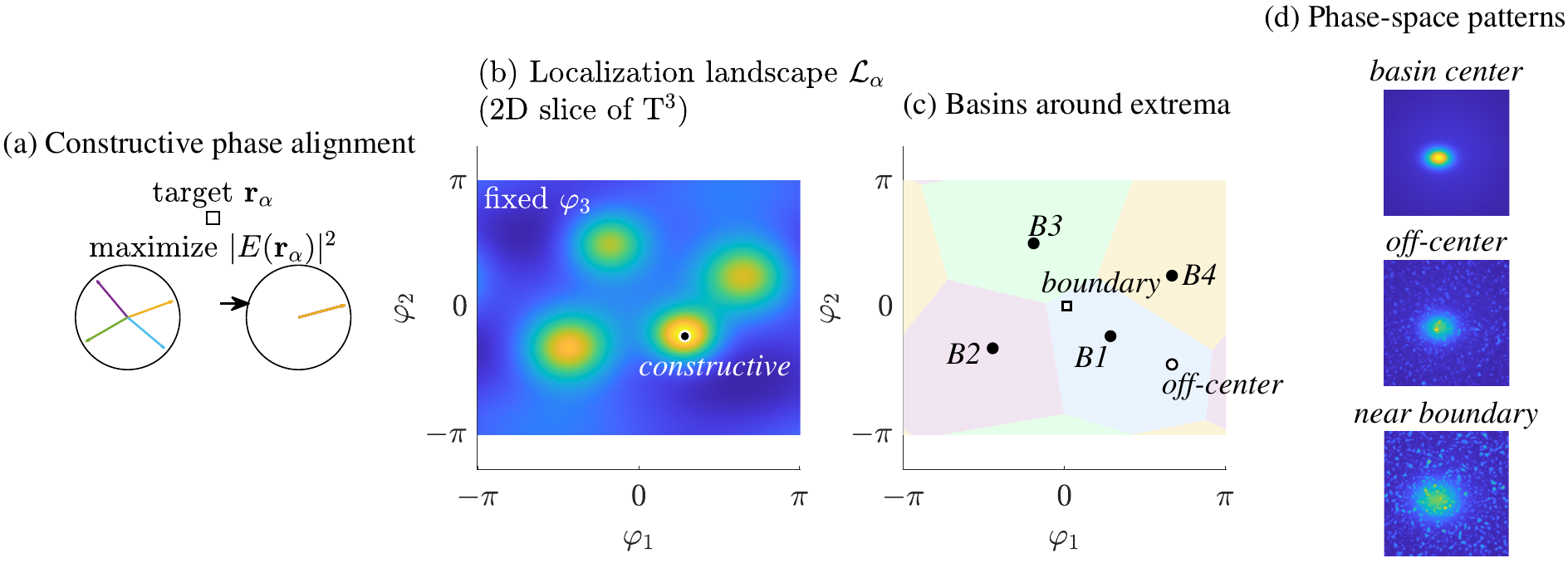}}
\caption{\textbf{Constructive phase solutions and localization basins under phase-only control.} \textbf{(a)} Schematic illustration of constructive phase alignment for localization at a target position $\mathbf{r}_\alpha$. Random relative phases produce weak interference at the target, whereas aligning the reduced phases according to $\varphi_{m-1}^{(\alpha)}=\arg[c_1u_1(\mathbf{r}_\alpha)]-\arg[c_mu_m(\mathbf{r}_\alpha)]$ (for $m=2,\ldots,N$) enforces constructive addition (up to a global phase) and maximizes the local intensity $|E(\mathbf{r}_\alpha)|^2$. \textbf{(b)} Localization landscape $\mathcal{L}(\boldsymbol{\varphi})$ evaluated over a two-dimensional slice of the reduced phase-space $\mathbb{T}^{N-1}$ (here $N=4$ with one reduced phase held fixed), revealing multiple isolated stationary points that act as centers of attraction. \textbf{(c)} Distinct spatially localized output fields associated with representative stationary points (basins) labeled $B_1$--$B_4$. Panel~(b) encodes the geometry of the phase-space control landscape, while panel~(c) encodes the identity of the localized states; matching labels, rather than spatial positions, establish the correspondence. \textbf{(d)} Representative output intensity patterns sampled across the phase-space slice, illustrating the continuous evolution of the spatial field between basins and the emergence of sharply localized states near basin centers.}
\label{fig3}
\end{figure*}
\medskip
\noindent\textbf{Constructive phase solutions and basin multiplicity.}
Localized states correspond to stationary points of the phase-space landscape with respect to the reduced phases. For quadratic localization measures, these arise when relative phases enforce globally consistent constructive interference, yielding isolated solutions on the compact manifold (see Supplementary Note~3). A transparent case is point localization, where phase alignment at the target maximizes local intensity under phase-only control.

Figure~\ref{fig3} illustrates the geometric consequences. Constructive conditions generate isolated stationary points in the reduced phase space, corresponding to discrete basin centers (Fig.~\ref{fig3}b). Each center defines a localized output with basin-dependent efficiency, consistent with experimentally observed states in photonic lanterns (Fig.~4 of Ref.~\cite{chandrasekharan2025}). Strong localization occurs only within finite neighborhoods of these centers, while configurations away from them interpolate continuously between output patterns (Fig.~\ref{fig3}c,d). The resulting multiplicity reflects compactness-imposed partitioning of phase space, independent of optimization or calibration. Phase conjugation identifies basin centers, whereas basin geometry—through curvature and stiffness—governs accessibility, robustness, and scaling under phase-only control.

\begin{figure*}[t]
\centering{\includegraphics[width=\linewidth]{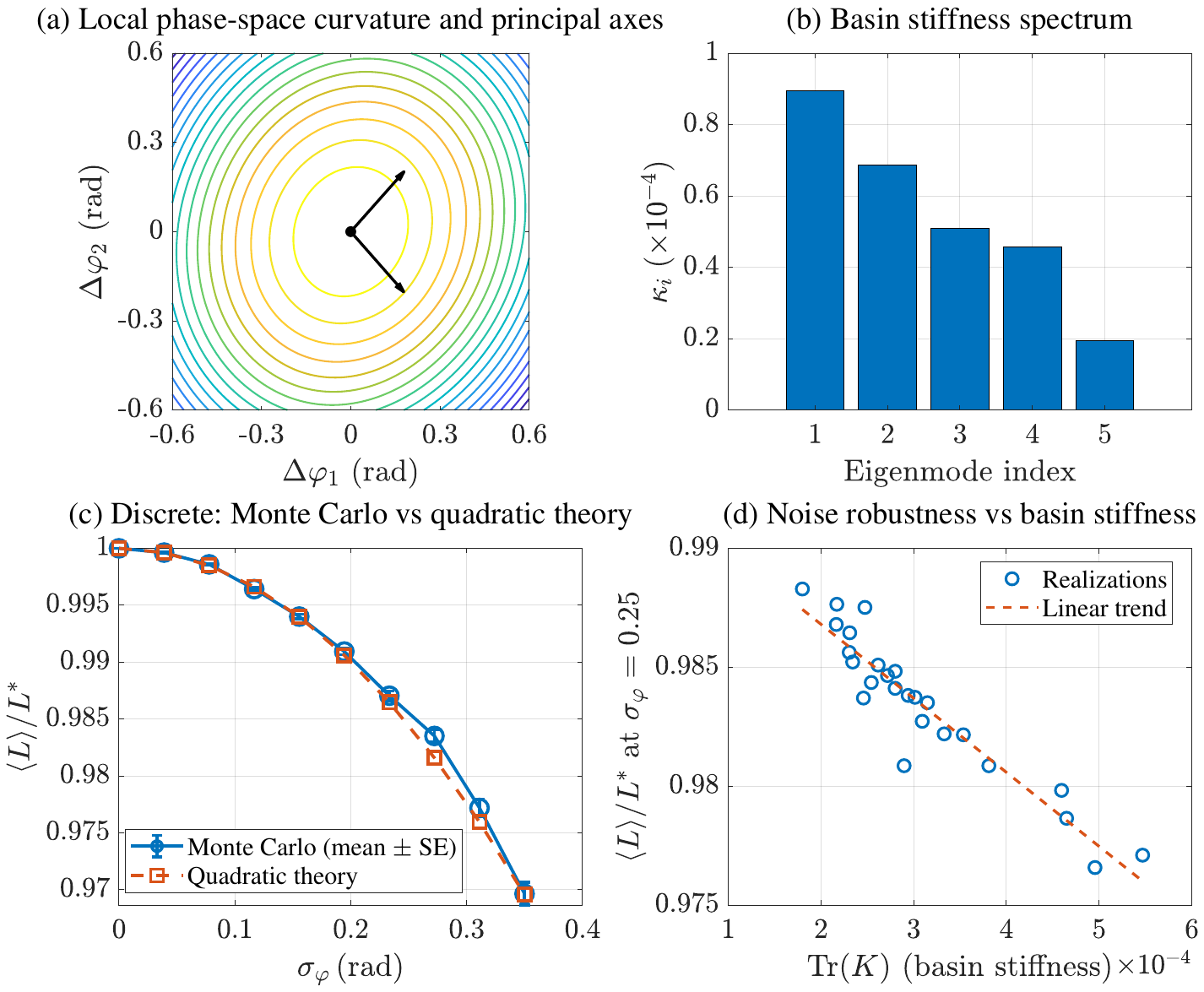}}
\caption{\textbf{Curvature and robustness of phase-space localization basins.} \textbf{(a)} Local quadratic approximation of the phase-space landscape around a stable basin center, showing contour lines of $\mathcal{L}(\boldsymbol{\varphi}^\star+\delta\boldsymbol{\varphi})$ in a two-dimensional phase slice and the principal curvature axes. \textbf{(b)} Basin stiffness spectrum given by the eigenvalues $\{\kappa_i\}$ of $K\equiv -H$, where $H$ is the Hessian at the basin center; in the numerical example shown here $N=6$, hence $N-1=5$ stiffness eigenvalues. \textbf{(c)} Localization degradation under discrete phase noise: Monte Carlo estimates of $\langle \mathcal{L}\rangle/L^\ast$ compared with the quadratic prediction $\langle \delta \mathcal{L}\rangle \approx -\tfrac{1}{2}\mathrm{Tr}(H\Sigma)$. $\mathrm{Tr}(H\Sigma)$ is extracted from small symmetric perturbations about the stationary configuration and remains valid within the quadratic regime (Supplementary Note~8). \textbf{(d)} Noise robustness versus basin stiffness, showing localization fidelity at a fixed RMS phase noise $\sigma_{\varphi}$ across different localization basins.}
\label{fig4}
\end{figure*}

\medskip
\noindent\textbf{Stability, robustness, and basin curvature.}
The utility of a localized state depends not only on the value of the localization functional at a basin center, but also on the geometry of the surrounding phase-space basin. As shown in Fig.~\ref{fig4}a, the landscape near a stable stationary point is generally anisotropic, with distinct principal curvature axes. Figure~\ref{fig4}b summarizes this structure through the basin stiffness spectrum, given by the eigenvalues of the Hessian \cite{strogatz:2000,MR0690288} at the stationary point.

The consequences of basin curvature are illustrated in Fig.~\ref{fig4}c,d. Localization degradation under discrete phase noise is accurately captured by a quadratic approximation based on local curvature (Fig.~\ref{fig4}c). Basin stiffness therefore provides a predictive geometric measure of noise robustness: basins with smaller curvature retain higher localization fidelity at fixed RMS phase noise (Fig.~\ref{fig4}d). Although curvature-based stability analysis is standard in unconstrained optimization, its role under phase-only control is distinct—curvature determines the finite volume and anisotropy of high-performance regions on the compact manifold, rather than convergence rates (see Supplementary Note~4).

\begin{figure*}[b]
\centering
\includegraphics[width=\linewidth]{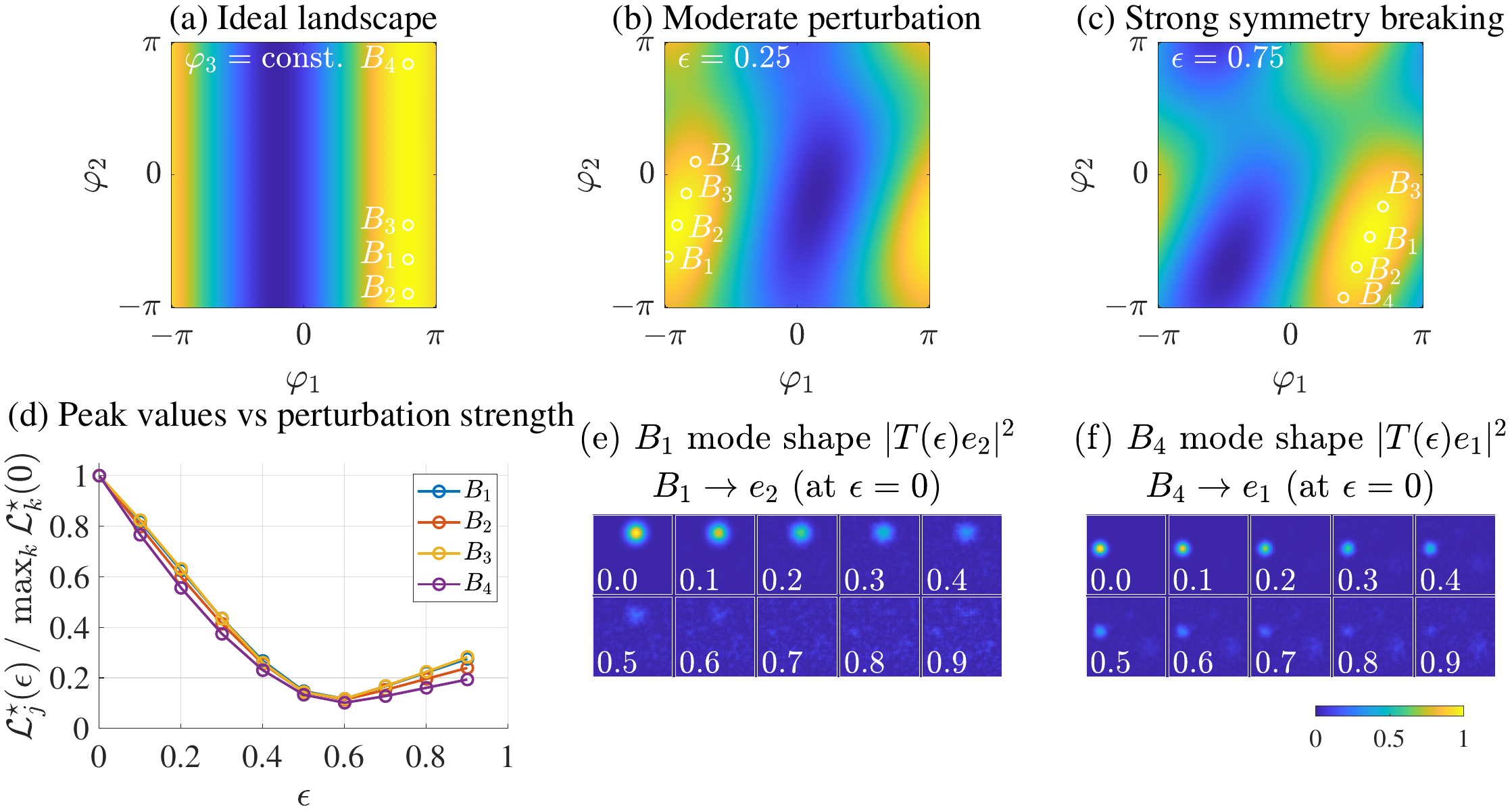}
\caption{\textbf{Imperfection-induced deformation of phase-space basins and associated mode shapes.}
\textbf{(a)} Ideal case: a two-dimensional slice of the localization functional
$\mathcal{L}_W(\varphi_2-\varphi_1,\varphi_3-\varphi_1)$ (with $\varphi_4$ held fixed for $N=4$),
exhibiting four basins of attraction with comparable depth.
\textbf{(b)} Moderate non-idealities, introduced through a continuous interpolation between an ideal
mode-selective mapping and a symmetry-breaking perturbation, displace basin centers and split
basin depths while preserving basin identity.
\textbf{(c)} Strong symmetry breaking can cause basins within a given phase-space slice to merge or lose distinct identity through bifurcation of extrema, signaling a crossover toward a leakage-dominated landscape.
\textbf{(d)} Normalized peak basin values
$\mathcal{L}_j^{\star}(\epsilon)/\max_k \mathcal{L}_k^{\star}(0)$ versus asymmetry strength
$\epsilon$ for basins $j=1,\ldots,4$, showing progressive degradation of localization, a minimum
at intermediate asymmetry, and a weak upturn at large $\epsilon$ due to accidental overlap of a
speckle-dominated field with the measurement region.
\textbf{(e,f)} Evolution of the effective mode shapes associated with basins $B_1$ and $B_4$,
respectively, shown as intensity maps $|T(\epsilon)e_{k^\ast}|^2$ of the corresponding
transmission columns, where the input index $k^\ast$ is determined by maximum overlap with the
basin-locked field at $\epsilon=0$. As asymmetry ($\epsilon$, indicated at the bottom) increases, the initially localized core develops a diffuse speckled halo, reflecting leakage into a high-dimensional background subspace rather than simple transfer of energy between basin centers. Panels (e,f) use a
power-law display (display gain=0.65) to enhance low-intensity speckle.
}
\label{fig5}
\end{figure*}
\section{Effects of imperfections and basis mixing}
\label{sec:imperfections}

Real multimode systems deviate from idealized modal descriptions due to mode-dependent loss, residual mixing, imperfect orthogonality, and leakage into uncontrolled channels. These imperfections deform the phase-space landscape and limit the robustness and fidelity of phase-induced localization.

\medskip
\noindent\textbf{Operator formulation of imperfections.}
To capture these effects, we represent the output field on a finite sampling basis as a complex vector $\mathbf{E}\in\mathbb{C}^{M}$ and describe the system by a linear operator $\mathbf{T}\in\mathbb{C}^{M\times N}$ acting on the phase-only modal coefficients. Quadratic localization measures then depend on an effective Gram matrix $\mathbf{G}_{\mathrm{eff}}=\mathbf{T}^\dagger\mathbf{W}\mathbf{T}$, which encodes non-idealities relevant to spatial concentration (see Supplementary Note~5). Imperfections therefore enter the phase-space landscape exclusively through $\mathbf{G}_{\mathrm{eff}}$, linking physical disorder to geometric deformation.

\medskip
\noindent\textbf{Deformation and loss of phase-space basins.}
Figure~\ref{fig5}a--c illustrates the impact of increasing imperfection strength. In the ideal case, symmetry of the effective modal ensemble produces multiple basins of comparable depth (Fig.~\ref{fig5}a). Moderate non-idealities displace basin centers and split basin depths while preserving basin identity as seen in Fig.~\ref{fig5}b. Under strong symmetry breaking, basins within a phase-space slice may merge or lose distinct identity through bifurcation of extrema (Fig.~\ref{fig5}c), signaling a crossover to a leakage-dominated landscape rather than disappearance of stationary solutions.

Figure~\ref{fig5}d quantifies the evolution of peak basin values with increasing imperfection. The initial decay reflects loss of constructive interference within the controlled (low-rank) subspace. The weak upturn at large imperfection does not indicate recovery of phase-only control, but results from accidental overlap between a speckle-dominated field and the measurement region, consistent with extreme-value statistics of high-dimensional random backgrounds.

\medskip
\noindent\textbf{Leakage-induced degradation of physical mode shapes.}
Imperfections act on two levels: they deform phase-space basin geometry and degrade the real-space structure of localized states. As shown in Fig.~\ref{fig5}e,f, initially well-localized cores persist under moderate imperfections but develop diffuse speckled halos as leakage into uncontrolled high-dimensional background modes increases. While basin structure determines which phase configurations remain accessible, localization quality is ultimately limited by distributed background leakage that cannot be compensated by phase-only control.

Together, these results reveal a fundamental limitation of phase-induced relocalization under constrained actuation. As imperfections grow, the effective rank and conditioning of $\mathbf{G}_{\mathrm{eff}}$ deteriorate, flattening the phase-space landscape and suppressing robust basins. Beyond this regime, fidelity is governed not by compact-manifold partitioning but by energy redistribution into high-dimensional speckle subspaces inaccessible to phase-only control.

\medskip
\noindent\textbf{Geometric scaling limit for compact phase control.}
Confinement of coherent dynamics to the compact manifold of relative phases has two structural consequences. Compactness and continuity of quadratic localization observables generically enforce a finite set of isolated stationary interference configurations, partitioning phase space into discrete basins. Simultaneously, dimensional mismatch between the $(N-1)$-dimensional controlled phase manifold and leakage-populated subspaces constrains achievable localization contrast through the single mismatch parameter $\Xi \equiv \ell (N-1)$.

\begin{figure*}[t]
\centering
\includegraphics[width=\linewidth]{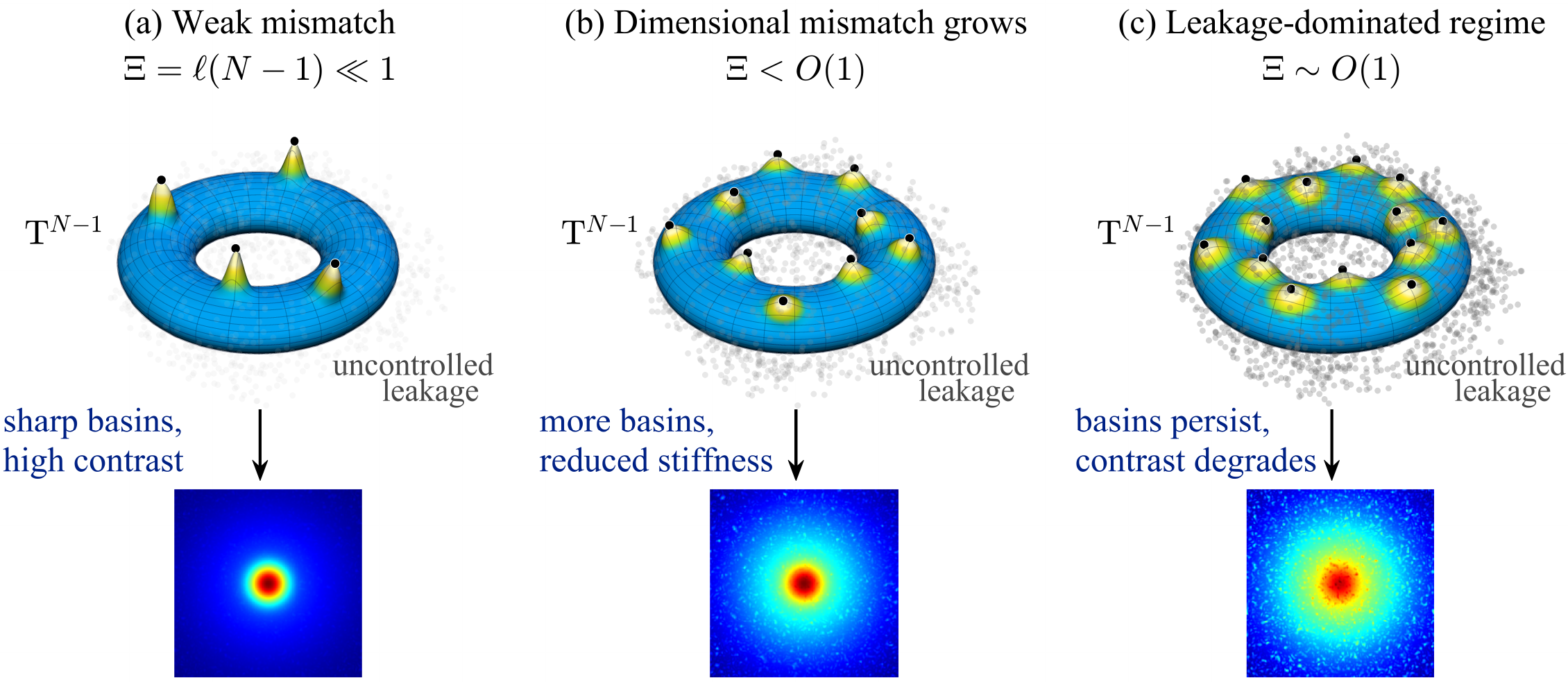}
\caption{\textbf{Geometric scaling limits of phase-only localization.}
Schematic illustration of the scaling law governed by the mismatch parameter $\Xi=\ell(N-1)$, where $\ell$ denotes leakage into uncontrolled degrees of freedom and $N$ is the number of controlled modes. The compact phase manifold $\mathrm{T}^{N-1}$ supports discrete stationary interference configurations (basins), while uncontrolled leakage populates a higher-dimensional background subspace. (\textbf{a}) Weak mismatch ($\Xi\ll1$): isolated sharp basins produce high-contrast localized states. (\textbf{b}) As dimensional mismatch increases, the number of accessible stationary states grows while basin stiffness decreases due to leakage-induced softening. (\textbf{c}) In the leakage-dominated regime ($\Xi\sim O(1)$), stationary configurations persist but achievable contrast and robustness degrade systematically as energy redistributes into uncontrolled subspaces. The lower panels illustrate the corresponding evolution from sharply localized outputs to diffuse, low-contrast localization.}
\label{fig6}
\end{figure*}

\medskip
\noindent\textbf{Geometric scaling law.}
Consider a coherent system with phase-only actuation on $N$ controlled modes and leakage fraction $\ell$, defined as the fractional weight of the field in uncontrolled degrees of freedom. Let $\eta$ denote the normalized contrast of a Hermitian quadratic observable. Stationary interference configurations generically persist on the compact phase manifold. Under weak leakage (i.e.\ $\ell \ll 1$), their maximal achievable contrast obeys
\[
\eta_{\max} \lesssim 1 - C\,\Xi,
\]
to leading order, where $\Xi \equiv \ell (N-1)$ and $C$ depends only on the geometric overlap between controlled and leakage-populated subspaces (schematically illustrated in Fig.~\ref{fig6}). For $\Xi \gtrsim O(1)$, the system enters a non-perturbative leakage regime in which stationary configurations persist by compactness while contrast, basin stiffness, and deterministic basin identity undergo order-unity degradation. Under partial control ($\ell > 0$), classical enhancement scaling with $N$ and compact-control scaling therefore become parametrically incompatible at large $N$ (see Supplementary Note~9). This implies that increasing mode count without controlling leakage will inevitably degrade performance, regardless of optimization strategy.

The result applies to any Hermitian quadratic observable defined on a compact phase manifold. Compactness guarantees persistence of stationary configurations, while dimensional mismatch with uncontrolled subspaces constrains their attainable contrast. More generally, the peak-contrast bound represents one observable manifestation of a broader compact-control deformation principle, under which contrast reduction, basin softening, and loss of deterministic basin identity arise as coupled geometric consequences (Supplementary Note~7). Experimentally, measuring basin stiffness versus $N$ at fixed imperfection should collapse when plotted against $\Xi$ (see Supplementary Note~9). Cross-platform realizations across classical and quantum coherent systems are detailed in Supplementary Notes~9 and~11.

\section{Photonic lantern realization of the geometric scaling constraint}
\label{sec:lantern}

The geometric scaling law follows from compact phase-only control and dimensional mismatch, independent of device architecture. As a concrete realization, we consider the mode-selective photonic lantern, which provides a controlled linear phase-to-field mapping where the compact $U(1)^{N-1}$ manifold and leakage-induced subspace coupling can be directly examined. The lantern does not generate the scaling law; rather, it renders the geometric mechanism explicit: constructive phase configurations persist as the number of controlled modes increases, while their achievable contrast degrades according to dimensional scaling.

\medskip
\noindent\textbf{Lantern as a phase-to-field linear map.}
A photonic lantern can be modeled as a linear map from $N$ single-mode inputs to $N$ output field degrees of freedom~\cite{Leon-Saval:10,Birks:15,Montoya:16}. After fixing the global phase, the phase-only input state is represented by a normalized vector $\mathbf{x}(\boldsymbol{\varphi}) \in \mathbb{C}^N$ whose components differ only by controllable relative phases. The output samples are $\mathbf{y}(\boldsymbol{\varphi}) = \mathbf{H}\mathbf{x}(\boldsymbol{\varphi})$, where $\mathbf{H}$ is a fixed transfer matrix. In an ideal lantern, $\mathbf{H}$ approximates a unitary equal-modulus demultiplexing transformation, enabling phase-only modulation to concentrate energy into a selected output channel.

\medskip
\noindent\textbf{Basin selection by constructive phase alignment.}
To select output channel $\alpha$, we use the target intensity $\mathcal{L}_\alpha = |y_\alpha|^2$ as the localization functional. Under phase-only control, it is maximized by constructive alignment of all modal contributions at the target, equivalent to a phase-conjugation condition on the input phases. This condition identifies the corresponding phase-space basin center and provides the lantern realization of the stationary interference solutions described in Section~\ref{sec:model} (see Supplementary Note~3).

\medskip
\noindent\textbf{Basin multiplicity versus basin quality.}
In an ideal lantern implementing balanced demultiplexing, there exist $N$ symmetry-related constructive phase configurations, each a stationary point of $\mathcal{L}_\alpha$ for a different target channel. Thus the number of accessible basins scales linearly with $N$, even in large systems. The relevant limitation, however, lies not in the \emph{existence} of stationary points but in their \emph{quality}. We quantify this by a normalized localization index $\eta_\alpha \in [0,1]$, measuring the fraction of total output power concentrated in the selected channel under constructive alignment. We refer to $\eta_\alpha$ as the basin quality metric for scaling, distinguishing it from peak basin values $L^\star$ used to characterize deformation at fixed $N$.
\begin{figure*}[b]
\centering
\includegraphics[width=\linewidth]{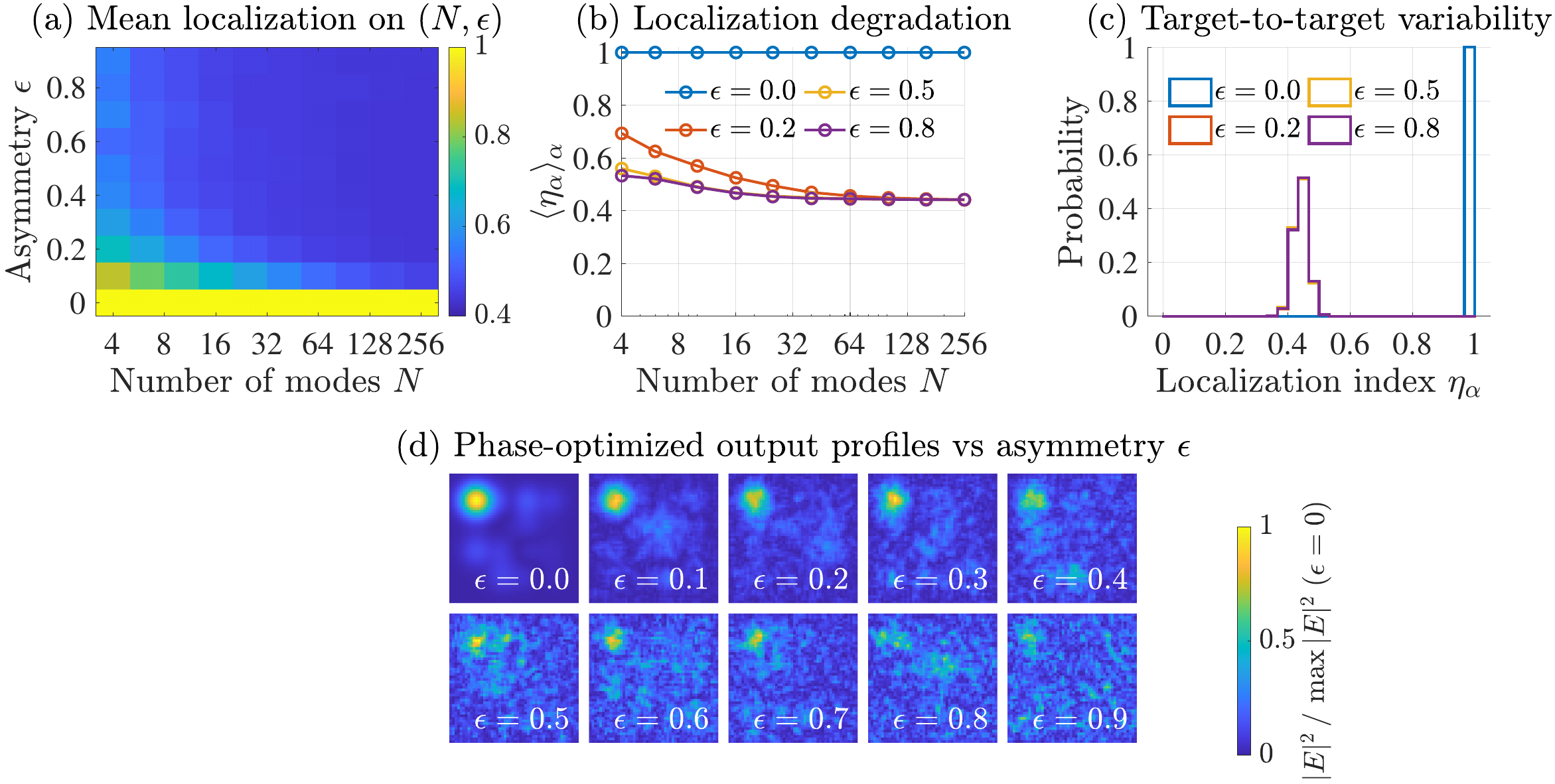}
\caption{\textbf{Scaling of basin quality in a photonic lantern under asymmetry.}
\textbf{(a)} Mean localization index $\langle \eta_\alpha \rangle$ as a function of the number of modes $N$ and asymmetry strength $\epsilon$. An ideal lantern ($\epsilon=0$) maintains near-unity localization for all $N$, whereas increasing asymmetry leads to a systematic degradation of basin quality that becomes more pronounced as $N$ grows. \textbf{(b)} Line cuts of panel (a) at selected values of $\epsilon$, highlighting the monotonic decrease of mean localization with increasing $N$ under fixed asymmetry. \textbf{(c)} Distribution of localization indices $\eta_\alpha$ across all target channels for $N=256$, illustrating the broadening and leftward shift of basin quality as asymmetry increases. At large $N$, the distributions collapse toward a characteristic mid-range localization level, reflecting saturation of effective mode mixing rather than further loss of phase-space stationary points. \textbf{(d)} Output intensity profiles under constructive phase alignment for a fixed $N=16$ lantern, showing the spatial structure of the best-case localized output under constructive phase alignment as $\epsilon$ increases from left to right. Together, these panels demonstrate that although the number of basins scales linearly with $N$ in principle, their practical utility is ultimately limited by imperfection-induced background leakage. The same geometric mechanism applies in coherent multimode systems whenever control is restricted to relative phases on a compact manifold (Supplementary Note~9).}
\label{fig7}
\end{figure*}
With leakage into uncontrolled degrees of freedom, constructive solutions persist but their achievable contrast degrades systematically with increasing modal dimensionality. The lantern therefore provides a direct illustration of the geometric scaling law (Section~\ref{sec:imperfections}): stationary configurations remain on the compact manifold while basin quality softens under dimensional mismatch.

Figure~\ref{fig7} shows this scaling. Panel (a) plots the mean localization index $\langle\eta_\alpha\rangle$ versus mode number $N$ and asymmetry $\epsilon$: an ideal lantern ($\epsilon=0$) maintains near-unity localization, whereas increasing asymmetry produces degradation that strengthens with $N$. Panel (b) shows the decay of mean localization with $N$ for selected $\epsilon$, and panel (c) shows the corresponding distributions at large $N$, which broaden and shift toward lower values as imperfections increase. The mean degradation therefore reflects broad basin softening rather than a small subset of poorly performing channels.

\medskip
\noindent\textbf{Scaling-induced degradation and leakage-dominated localization.}
The physical manifestation of the scaling constraint is illustrated in Fig.~\ref{fig7}d for a representative lantern ($N = 16$). An ideal system yields a sharply localized mode, whereas increasing asymmetry preserves the core but generates a stronger speckled background, reflecting geometric degradation: imperfect control conditions the effective Gram operator $G_{\mathrm{eff}}$, producing distributed leakage into uncontrolled degrees of freedom without eliminating stationary configurations (Supplementary Note~6). As a result, high-contrast localization is limited by background intensity inaccessible to phase-only control. In the large-$N$ limit, performance is therefore set by basin \emph{quality}, not basin \emph{count}: multiple basin centers persist, but their contrast and stiffness degrade systematically with system size.

\begin{figure*}[b]
\centering
\includegraphics[width=\linewidth]{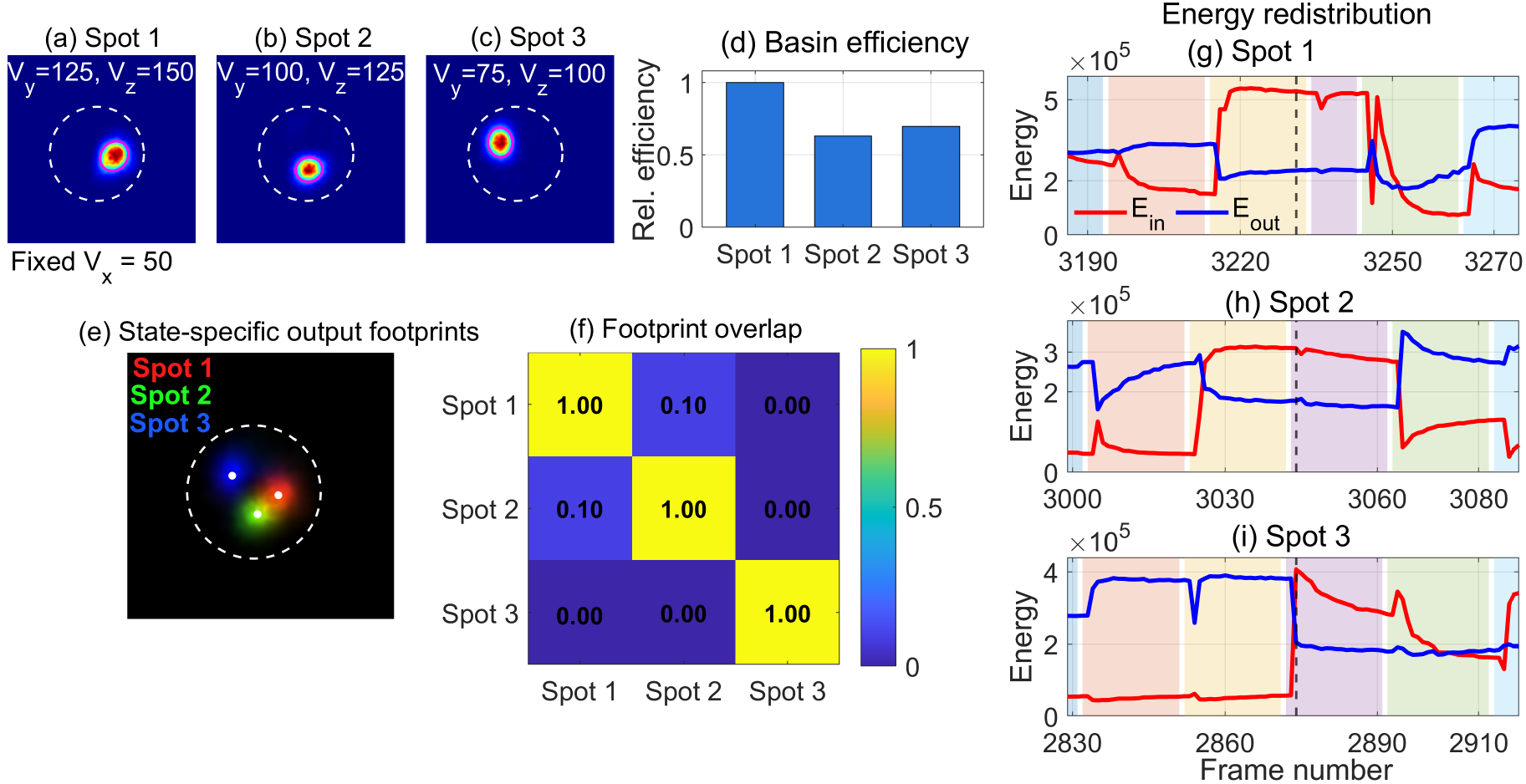}
\caption{\textbf{Experimentally identified operating regimes for controllable localization in a multimode system under phase-only control.}
\textbf{(a--c)} Representative localized output states (Spot 1--3) obtained from frames sampled during on-the-fly voltage scans, at selected operating points with fixed $V_x=50$ and varying $(V_y, V_z)$. Each panel shows the measured intensity distribution within the core region (dashed circle), highlighting distinct spatial localization patterns associated with different voltage configurations.
\textbf{(d)} Relative basin efficiency $E_{\mathrm{rel}}$, defined as the maximum localized energy within the selected operating segment, normalized to the highest-performing basin across all sampled states. This metric compares the best achievable localization efficiency of each experimentally accessible basin.
\textbf{(e)} State-specific sampled-flux landscapes in a common field of view. Colored maps (red: Spot 1, green: Spot 2, blue: Spot 3) are obtained by averaging normalized intensity distributions over the frames belonging to the selected plateau segment (see (g–i)), capturing the spatial footprint of each operating basin under coarse voltage stepping. White markers indicate the centroid of each localized state. Partial overlap and asymmetry arise from the proximity of basins in control space and non-idealities in the system, including fabrication imperfections and residual mode coupling, which can enable transition pathways between states.
\textbf{(f)} Footprint overlap matrix computed using the Jaccard index (intersection over union) between fixed-level footprints. The low off-diagonal values indicate that the sampled operating states are largely distinct, with limited but non-zero overlap between certain basins.
\textbf{(g--i)} Energy redistribution across sampled voltage configurations for each state. The target-segment operating point is marked by a dashed vertical line. The traces show energy inside ($E_{\mathrm{in}}$) and outside ($E_{\mathrm{out}}$) the core region as control voltages are varied. Shaded regions denote contiguous voltage segments. These plots reveal plateau-like regions of stable operation and illustrate how different states trade off stability, efficiency, and robustness under coarse voltage stepping.}
\label{fig8}
\end{figure*}

\section{Experimental identification of robust multimode operating regimes via calibration-free phase scans}

Phase-space basin geometry provides a direct, calibration-free route for identifying stable operating states in multimode systems under phase-only control. We demonstrate this in a three-mode photonic lantern platform operating at telecom wavelengths~\cite{chandrasekharan2025}, where voltage-driven actuation serves as a proxy for relative phase control. A coarse voltage scan samples the compact phase manifold, generating interference states in which localized output spots appear at discrete phase configurations (see Methods and Supplementary Note~12). These states correspond to distinct basins of the phase landscape, with intensity reflecting basin efficiency and spatial confinement indicating robustness.

After fixing the reference voltage $V_x$, one redundant phase coordinate is removed, leaving a two-dimensional experimentally accessible control space sampled through $V_y$ and $V_z$. Continuous scans of $V_y$ and $V_z$ map the phase-space structure through frame-resolved intensity measurements (see Supplementary Figure~S5), revealing transitions between localized and intermediate interference states. Distinct localized states (Fig.~\ref{fig8}a--c) correspond to separate basins with different efficiencies (Fig.~\ref{fig8}d). Their spatial footprints and overlaps (Fig.~\ref{fig8}e,f) indicate varying basin isolation, consistent with imperfection-induced softening of basin boundaries. Temporal sampling provides a direct measure of stability: stable basins appear as plateau regions in energy traces (Fig.~\ref{fig8}g--i), while unstable configurations exhibit fluctuations. Among the observed states, Spot~1 combines the highest efficiency with strong temporal stability, identifying it as the most robust operating basin under the present conditions.

These results establish that phase-space basin geometry can be accessed experimentally without transmission-matrix reconstruction or iterative optimization. Coarse phase scans directly map accessible operating regimes, quantifying basin efficiency, isolation, and stability while identifying robust, high-efficiency states without system inversion or calibration. This is particularly relevant for practical multimode platforms subject to environmental drift and fabrication imperfections, where maintaining a single optimized configuration is not feasible, including applications in mode-division multiplexing, beam delivery, and adaptive coupling.

\section*{Discussion}

Compact phase-space geometry defines a distinct regime of coherent control under phase-only constraints, with direct implications for high-dimensional photonic systems based on wavefront shaping, multimode communications, coherent combining, and adaptive beam delivery. Confinement to the compact manifold of relative phases yields isolated stationary interference configurations, while mismatch with uncontrolled subspaces limits contrast, making the absence of continuously steerable localization geometric rather than algorithmic. Weak mismatch leads to gradual contrast reduction and basin softening, whereas strong mismatch drives a crossover to leakage-dominated behavior in which stationary configurations persist but robustness degrades (Supplementary Note~9).

This contrasts with unconstrained wavefront-shaping and eigenchannel approaches, where enhancement limits follow from the singular-value structure of the full transmission operator~\cite{PhysRevLett.104.100601}. Under phase-only actuation, constructive interference selects discrete basin centers, with robustness governed by basin curvature rather than full-field controllability, while dissipation and non-Hermitian effects act as smooth perturbations that soften basin boundaries without eliminating stationary configurations~\cite{doi:10.1126/science.aar7709,2018NatPh,Vellekoop:07}.

This structure defines the limits of low-dimensional phase control: basin curvature sets sensitivity to noise and resolution, providing an experimentally accessible robustness metric complementary to classical enhancement bounds~\cite{shekel2025f}. Imperfections—loss, residual mixing, and leakage—act as smooth perturbations that soften basin structure while preserving stationary states, with degradation reflecting redistribution to uncontrolled subspaces (Supplementary Note~6). Extensions to nonlinear and time-dependent regimes may yield evolving interference landscapes or multistability~\cite{10.106311,articlewrgt}, while quantum coherent systems exhibit analogous geometric constraints under phase-restricted control~\cite{Chakrabarti01102007}.

Photonic lanterns provide one realization of these principles~\cite{Montoya:16,Lin:25,Becerra-Deana:25}, but the framework is general. Any linear coherent system with phase-constrained actuation admits an effective Gram-operator description (Supplementary Note~5), so basin geometry and stiffness spectra provide broadly applicable diagnostics of robustness, independent of device architecture, and can be experimentally inferred from quadratic localization degradation under controlled phase perturbations without transmission-matrix reconstruction (Supplementary Note~8). Cross-platform realizations are detailed in Supplementary Note~9, and comparisons with existing control frameworks are given in Supplementary Note~10.

The experimental results show that coarse phase-only scans directly reveal the accessible phase space, with stable basins appearing as plateau regions in the measured response. This enables rapid identification of stable, high-efficiency operating regimes directly from measurements, without transmission-matrix reconstruction or adaptive optimization. Such calibration-free operation is particularly advantageous in practical multimode platforms subject to environmental drift and fabrication imperfections, where maintaining a single optimized configuration is not feasible. Compact phase-space geometry therefore provides a direct and experimentally accessible framework for identifying and exploiting robust operating regimes in complex coherent systems under constrained control.

\section*{Methods}

Throughout, $\epsilon$ denotes the strength of fabrication-induced asymmetry (imperfection). In the operator model described in Supplementary Note~5, increasing $\epsilon$ induces leakage into uncontrolled degrees of freedom, quantified by a leakage fraction $\ell(\epsilon)$. In Supplementary Note~6, we introduce a separate phenomenological leakage parameter $\gamma$ only for the probabilistic basin-softening schematic.

\noindent\textbf{Numerical evaluation of phase-space landscapes.}

Phase-space landscapes shown in Figs.~\ref{fig2}, \ref{fig3}, and \ref{fig4} were obtained by sampling reduced phase coordinates uniformly over two-dimensional slices of $\mathbb{T}^{N-1}$ while holding the remaining phases fixed. Localization functionals were evaluated directly from the modal superposition in Eq.~(\ref{eq:modal_expansion}), using the same modal coefficients and effective mode profiles as in the Supplementary simulations. Monte Carlo phase-noise simulations in Fig.~\ref{fig4} were performed by drawing random phase perturbations from zero-mean Gaussian distributions with prescribed covariance and averaging over $10^4$ realizations.

\noindent\textbf{Numerical modeling of imperfections.}

Imperfections were introduced using the operator model described in Supplementary Note~5, by continuously interpolating between an ideal mode-selective mapping and a symmetry-breaking perturbation. An imperfection parameter controls the relative weight of the perturbation and thus the degree of leakage into uncontrolled degrees of freedom. For each value of the imperfection strength, phase-space landscapes were evaluated by sampling reduced phase coordinates on two-dimensional slices of $\mathbb{T}^{N-1}$ while holding the remaining phases fixed. Peak basin values were extracted by local maximization of the localization functional within each basin.

\noindent\textbf{Lantern scaling simulations.}

Lantern transfer matrices were constructed using the operator model described in Supplementary Notes~5–6, by interpolating between an ideal balanced demultiplexing map and a dense symmetry-breaking perturbation. An asymmetry parameter $\epsilon$ controls the relative weight of the perturbation and thus the degree of leakage into uncontrolled degrees of freedom. For each pair of $(N,\epsilon)$, constructive phase solutions were computed analytically, and the localization index $\eta_\alpha$ was evaluated for all target channels. Ensemble averages and distributions were obtained by sampling over independent realizations of the perturbation operator.

\noindent\textbf{Experimental platform and measurement protocol.}

Experiments were performed using a commercial three-mode graded-index photonic lantern operating at 1550\,nm and supporting the LP$_{01}$, LP$_{11a}$, and LP$_{11b}$ modes, consistent with the platform previously reported in Ref.~\cite{chandrasekharan2025}. The measurements are analyzed here within the phase-space basin framework to identify robust operating regimes under constrained phase-only control. Coherent mode recombination was implemented in an all-fiber, polarization-reciprocal architecture using Faraday-reflector feedback, ensuring stable phase control without polarization alignment. Relative modal phases were controlled using piezoelectric actuators embedded in fiber delay lines at each lantern output, where applied voltages induce phase shifts via controlled optical path-length variations (see Figure S4, Supplementary Note 12). 

Before phase scanning, the optical path lengths were coarsely equalized using pulsed-laser time-of-flight measurements to ensure modal overlap within the laser coherence length. Phase-space sampling was then performed by continuously scanning the control voltages while recording the multimode output intensity using an InGaAs camera at $\sim$5\,fps, with voltages updated every $\sim$2\,s. This temporal oversampling yields $\sim$10 frames per operating point, enabling direct identification of stable basins through plateau behaviour in the energy traces. Frame-wise integrated intensity within the multimode core provides a synchronous measure of interference efficiency, while aggregated intensity maps over stable segments define the state-specific spatial footprints used to extract centroid positions, overlaps, and basin efficiency metrics.

\section*{Additional Information}

See the Supplementary Material for more information.\\

\section*{Acknowledgments}

The author thanks Innovate UK (10002685); Royal Academy of Engineering (RF\textbackslash201718\textbackslash1746); Engineering and Physical Sciences Research Council (EP/T001011/1); and Innovate UK (10005967) for funding support. The author thanks Dr Ross Donaldson (Heriot-Watt University) for providing access to the experimental platform and for permission to use the dataset employed in this study.

\section*{Declarations}
\subsection*{Conflict of Interest}
The author declares no conflict of interest.
\section*{Data availability}
The data that support the findings of this study are openly available in the Heriot-Watt University PURE research data management system at \url{https://doi.org/10.17861/46fba7ee-0a47-4b8b-a5d9-719ebfbb9325}

\section*{Code availability}
MATLAB codes used to generate the simulated figures are available at
\href{https://github.com/harikumarkvm/Phase-space-basin-geometry-a-framework-for-robust-phase-only-multimode-control.git}{GitHub/phase\_base\_geometry\_framework}.

%\input{References.tex} 
%this is to edit the letter e in the citaion. download .bbl output..save as.tex,open and insert the letter then commend bibliography below, and uncoment this....

%\bibliographystyle{sn-mathphys-num}
\providecommand{\noopsort}[1]{}\providecommand{\singleletter}[1]{#1}%

\clearpage


\section*{Supplementary material}

## Supplementary Note 1: Universal model for phase-controlled multimode fields

We consider a general coherent multimode system whose output field can be written as a finite superposition of spatial modes [1],

$$E(\mathbf{r};\boldsymbol{\phi}) = \sum_{m=1}^{N} c_m\, u_m(\mathbf{r})\, e^{i\phi_m}, \tag{1}$$

where $u_m(\mathbf{r})$ denote mode functions spanning the relevant modal subspace, $c_m$ are complex-valued modal amplitudes encoding fixed excitation weights and system-dependent transfer coefficients, and $\phi_m$ are externally controllable phases.

Equation (1) is completely general for linear, coherent systems and does not require orthogonality, equal normalization, or lossless propagation of the modes. Any static mode coupling, amplitude imbalance, or attenuation is absorbed into the coefficients $c_m$ and effective mode profiles $u_m(\mathbf{r})$, provided that coherence is preserved.

Because multiplication of all terms in Eq. (1) by a common phase factor leaves all physical observables invariant, the system exhibits a global phase (gauge) redundancy. This reduction from full phases to relative phases (and the resulting $(N-1)$-torus of control parameters) is standard in coherent wave superposition and interference theory [2, 3]. As a consequence, only $N-1$ independent phase degrees of freedom exist. Fixing the gauge by choosing one reference phase (without loss of generality $\phi_1 = 0$), we define the reduced phase coordinates

$$\boldsymbol{\varphi} = (\varphi_1, \ldots, \varphi_{N-1}), \qquad \varphi_j \equiv \phi_{j+1} - \phi_1. \tag{2}$$

With this convention, the modal expansion becomes

$$E(\mathbf{r};\boldsymbol{\varphi}) = c_1 u_1(\mathbf{r}) + \sum_{m=2}^{N} c_m\, u_m(\mathbf{r})\, e^{i\varphi_{m-1}}. \tag{3}$$

The accessible phase space is therefore the compact manifold $\mathbb{T}^{N-1}$, the $(N-1)$-dimensional torus arising from the $2\pi$-periodicity of each relative phase. Each point on $\mathbb{T}^{N-1}$ corresponds uniquely to a physically realizable output field configuration, and continuous variation of $\boldsymbol{\varphi}$ induces continuous deformation of the resulting multimode interference pattern.

Within this formulation, the reduced phases $\boldsymbol{\varphi}$ constitute the only controllable parameters, while the modal amplitudes $c_m$, mode profiles $u_m(\mathbf{r})$, and any static mode mixing or loss

are treated as fixed but generally unknown system characteristics. This separation between controlled and uncontrolled parameters reflects typical experimental conditions and underlies the generality of the phase-space framework developed in the main draft.

## Supplementary Note 2: Localization functionals and basin definitions

To characterize spatial localization independently of any specific physical platform, we introduce localization functionals that assign a scalar measure of spatial concentration to a given output field configuration. For a field $E(\mathbf{r};\boldsymbol{\varphi})$ defined by the phase-controlled modal superposition in Eq. (1), a localization functional $\mathcal{L}$ is defined as a real-valued mapping

$$\mathcal{L} : \mathbb{T}^{N-1} \to \mathbb{R}, \tag{4}$$

whose value depends on the reduced phase vector $\boldsymbol{\varphi}$ only through the resulting spatial intensity distribution $|E(\mathbf{r};\boldsymbol{\varphi})|^2$. The only assumptions placed on $\mathcal{L}$ are invariance under global phase transformations, continuity on $\mathbb{T}^{N-1}$, and sensitivity to spatial concentration of the field.

Evaluating $\mathcal{L}$ over the phase space defines a phase-space landscape $\mathcal{L}(\boldsymbol{\varphi})$. Because $\mathbb{T}^{N-1}$ is compact and $\mathcal{L}$ is continuous, this landscape necessarily contains extrema. Localized states correspond to stable local maxima of $\mathcal{L}$, i.e., stationary points with negative-definite Hessian.

To make the notion of a basin precise, we distinguish two related but conceptually distinct sets.

**Basin of attraction (dynamical definition).** Gradient flows and their associated basins of attraction are standard objects in dynamical systems and nonlinear stability theory [4]. We therefore consider the gradient-ascent flow on the torus,

$$\frac{\mathrm{d}\boldsymbol{\varphi}}{\mathrm{d}t} = \nabla_{\boldsymbol{\varphi}} \mathcal{L}(\boldsymbol{\varphi}), \tag{5}$$

interpreted with $2\pi$-periodic coordinates. For a stable local maximum $\boldsymbol{\varphi}^*$, its basin of attraction is defined as

$$\mathcal{A}(\boldsymbol{\varphi}^*) \equiv \left\{ \boldsymbol{\varphi} \in \mathbb{T}^{N-1} : \lim_{t\to\infty} \boldsymbol{\varphi}(t; \boldsymbol{\varphi}(0) = \boldsymbol{\varphi}) = \boldsymbol{\varphi}^* \right\}. \tag{6}$$

This definition formalizes the partition of phase space into regions that converge to distinct maxima under local ascent dynamics.

**Robustness neighborhood (level-set definition).** To quantify tolerance to finite phase perturbations independently of any specific dynamical rule, we define a purely geometric high-performance neighborhood around $\boldsymbol{\varphi}^*$ as the level set

$$\mathcal{R}_\delta(\boldsymbol{\varphi}^*) \equiv \left\{ \boldsymbol{\varphi} \in \mathbb{T}^{N-1} : \ \mathcal{L}(\boldsymbol{\varphi}) \geq \mathcal{L}(\boldsymbol{\varphi}^*) - \delta \right\}, \tag{7}$$

where $\delta > 0$ specifies an operational tolerance. Throughout the main draft, the term "basin" refers to the attraction basin $\mathcal{A}(\boldsymbol{\varphi}^*)$, while $\mathcal{R}_\delta(\boldsymbol{\varphi}^*)$ provides a complementary measure of robustness within that basin.

Related level-set notions are widely used to quantify robustness under finite perturbations in optimization and control landscapes [5].

**Existence and generic multiplicity of basin centers.** The compactness of the reduced phase space $\mathbb{T}^{N-1}$ and continuity of $\mathcal{L}$ guarantee the existence of at least one extremum of the localization landscape. Beyond this minimal statement, physically relevant phase-only wave superpositions generically produce *multiple isolated stationary points.* This follows from the non-separable, interference-driven dependence of $\mathcal{L}(\boldsymbol{\varphi})$ on the relative phases when multiple spatial modes contribute with distinct spatial profiles. Exact degeneracies or continuous families of maximizers require fine-tuned symmetries and are therefore nongeneric: arbitrarily small perturbations to modal amplitudes, spatial overlaps, or the target window lift such degeneracies and yield isolated critical points. Throughout this work, these isolated local maxima serve as basin centers, whose local curvature and global arrangement on $\mathbb{T}^{N-1}$ govern robustness, accessibility, and scaling behavior.

## Supplementary Note 3: Constructive phase solutions and basin multiplicity

Localized states correspond to stationary points of the localization functional $\mathcal{L}(\boldsymbol{\varphi})$ on the reduced phase space $\mathbb{T}^{N-1}$,

$$\nabla_{\boldsymbol{\varphi}} \mathcal{L}(\boldsymbol{\varphi}) = \mathbf{0}. \tag{8}$$

For quadratic localization functionals of the form

$$\mathcal{L}_W(\boldsymbol{\varphi}) = \int W(\mathbf{r})\, |E(\mathbf{r};\boldsymbol{\varphi})|^2\, \mathrm{d}\mathbf{r}, \tag{9}$$

where $W(\mathbf{r}) \geq 0$ defines a Hermitian positive semidefinite weighting operator (e.g. a spatial window or projection), substituting the gauge-fixed modal expansion yields

$$\mathcal{L}_W(\boldsymbol{\varphi}) = \sum_{m,n=1}^{N} c_m c_n^*\, e^{i(\tilde{\phi}_m(\boldsymbol{\varphi})-\tilde{\phi}_n(\boldsymbol{\varphi}))}\, G_{mn}, \qquad G_{mn} = \int W(\mathbf{r})\, u_m(\mathbf{r}) u_n^*(\mathbf{r})\, \mathrm{d}\mathbf{r}. \tag{10}$$

with lifted phases defined by $\tilde{\phi}_1 = 0$ and $\tilde{\phi}_m = \varphi_{m-1}$ for $m \geq 2$. Weighted Gram matrices provide a standard quadratic-form representation of interference and energy concentration measures in finite-dimensional modal subspaces [6, 7].

Differentiating with respect to the reduced phases gives the explicit stationarity condition

$$\frac{\partial \mathcal{L}_W}{\partial \varphi_j} = 2\,\Im\left[c_{j+1} e^{i\varphi_j} \sum_{n=1}^{N} c_n^* e^{-i\tilde{\phi}_n(\boldsymbol{\varphi})}\, G_{(j+1)n}\right] = 0, \qquad j = 1, \ldots, N-1. \tag{11}$$

Stationary solutions therefore correspond to reduced-phase configurations that eliminate the net out-of-phase interference contribution of each controlled modal component within the weighted region.

For point localization at $\mathbf{r}_\alpha$, where $W(\mathbf{r}) = \delta(\mathbf{r} - \mathbf{r}_\alpha)$, the localization functional reduces to

$$\mathcal{L}_\alpha(\boldsymbol{\varphi}) = |E(\mathbf{r}_\alpha;\boldsymbol{\varphi})|^2. \tag{12}$$

A sufficient constructive reduced-phase solution under the gauge choice $\tilde{\phi}_1 = 0$ is

$$\varphi_{m-1}^{(\alpha)} = \arg[c_1 u_1(\mathbf{r}_\alpha)] - \arg[c_m u_m(\mathbf{r}_\alpha)] \mod 2\pi, \qquad m = 2, \ldots, N, \tag{13}$$

provided that $c_m u_m(\mathbf{r}_\alpha) \neq 0$ for all contributing modes. This phase choice aligns all contributing phasors at $\mathbf{r}_\alpha$ and therefore maximizes the point intensity

$$L_\alpha(\phi) = |E(\mathbf{r}_\alpha;\phi)|^2, \tag{14}$$

under phase-only control. In general, however, it does not necessarily maximize a finite-window localization functional $L_W$.

These constructive solutions define centers of attraction in phase space and are generally not

unique due to phase periodicity, spatial symmetries of the modal ensemble, and degeneracies in modal overlaps. Each stationary solution $\boldsymbol{\varphi}^\star$ defines a basin of attraction $\mathcal{A}(\boldsymbol{\varphi}^\star) \subset \mathbb{T}^{N-1}$ under the gradient-ascent dynamics defined in Note 2.

## Supplementary Note 4: Stability, robustness, and curvature of phase basins

To quantify the local stability of a localization basin centered at a stationary point $\boldsymbol{\varphi}^\star$, we expand the localization functional $\mathcal{L}$ to second order in phase perturbations $\delta\boldsymbol{\varphi}$:

$$\mathcal{L}(\boldsymbol{\varphi}^\star + \delta\boldsymbol{\varphi}) = \mathcal{L}(\boldsymbol{\varphi}^\star) + \nabla_{\boldsymbol{\varphi}}\mathcal{L}(\boldsymbol{\varphi}^\star)^\mathsf{T}\delta\boldsymbol{\varphi} + \frac{1}{2}\delta\boldsymbol{\varphi}^\mathsf{T} H \delta\boldsymbol{\varphi} + \mathcal{O}(|\delta\boldsymbol{\varphi}|^3), \tag{15}$$

where the Hessian matrix is

$$H_{ij} = \left.\frac{\partial^2 \mathcal{L}}{\partial\varphi_i \partial\varphi_j}\right|_{\boldsymbol{\varphi}^\star}. \tag{16}$$

At a stationary point, $\nabla_{\boldsymbol{\varphi}}\mathcal{L}(\boldsymbol{\varphi}^\star) = \mathbf{0}$, so the leading variation is quadratic. Local stability requires $H$ to be negative definite. Hessian-based curvature criteria for local stability and perturbation sensitivity are standard in nonlinear dynamics and optimization [4, 5].

Defining the stiffness matrix $K \equiv -H$, we diagonalize $K = Q\,\mathrm{diag}(\kappa_1, \ldots, \kappa_{N-1})Q^\mathsf{T}$ with $\kappa_i > 0$. For random phase perturbations with covariance $\Sigma = \langle \delta\boldsymbol{\varphi}\, \delta\boldsymbol{\varphi}^\mathsf{T} \rangle$, the expected change in localization is

$$\langle \delta\mathcal{L} \rangle \approx \frac{1}{2}\,\mathrm{Tr}(H\,\Sigma) = -\frac{1}{2}\,\mathrm{Tr}(K\,\Sigma)\,. \tag{17}$$

For fixed noise statistics (fixed $\Sigma$), this expression shows that differences in basin stiffness directly translate into differences in localization fidelity across distinct basins.

For perturbations bounded in Euclidean norm, $\|\delta\boldsymbol{\varphi}\|_2 \le \rho$, the quadratic approximation yields deterministic bounds on localization degradation,

$$0 \le \mathcal{L}(\boldsymbol{\varphi}^\star) - \mathcal{L}(\boldsymbol{\varphi}^\star + \delta\boldsymbol{\varphi}) \approx \frac{1}{2}\,\delta\boldsymbol{\varphi}^\mathsf{T} K\,\delta\boldsymbol{\varphi} \le \frac{1}{2}\,\kappa_{\max}\,\rho^2, \tag{18}$$

where $\kappa_{\max} = \max_i \kappa_i$. Here $\rho$ denotes the Euclidean-norm bound on phase perturbations (a phase-noise amplitude), introduced to avoid overloading the imperfection parameter $\epsilon$ used in the main text.

More generally, since $K$ is positive definite,

$$\frac{1}{2}\,\kappa_{\min}\,\|\delta\boldsymbol{\varphi}\|_2^2 \le \mathcal{L}(\boldsymbol{\varphi}^\star) - \mathcal{L}(\boldsymbol{\varphi}^\star + \delta\boldsymbol{\varphi}) \approx \frac{1}{2}\,\delta\boldsymbol{\varphi}^\mathsf{T} K\,\delta\boldsymbol{\varphi}, \tag{19}$$

where equality with the lower bound is attained for perturbations aligned with the weakest-curvature eigenvector.

These relations imply that, for a fixed perturbation amplitude, basins with larger curvature experience larger localization loss, while basins with smaller curvature retain higher fidelity. Together, these bounds quantify how basin curvature controls robustness and determine the local geometry of the robustness region $R_\delta(\varphi^\star)$ defined in Supplementary Note 2 [Eq. (7)], with $\delta$ set by the tolerated localization loss induced by perturbations of amplitude $\rho$.

The quadratic degradation of the localization functional under small phase perturbations provides direct access to the local Hessian and basin stiffness spectrum. The full experimental protocol, including phase sampling strategy and statistical estimation procedures, is provided in Supplementary Note 8.

## Supplementary Note 5: Operator model for imperfections and basis mixing

To capture both controlled and uncontrolled degrees of freedom [8, 9], it is useful to decompose the system operator as

$$\mathbf{T}(\epsilon) = \sqrt{1-\ell(\epsilon)}\,\mathbf{U}\mathbf{A}(\epsilon) + \sqrt{\ell(\epsilon)}\,\mathbf{S}\mathbf{B}(\epsilon), \tag{20}$$

where $\mathbf{U} \in \mathbb{C}^{M\times N}$ spans a designed low-rank subspace supporting localized mode cores, $\mathbf{A}(\epsilon)$ encodes controlled mixing and mode-dependent loss, and $\mathbf{S} \in \mathbb{C}^{M\times K_s}$ spans a speckle-like background subspace with $K_s \gg N$ [7]. The leakage parameter $\ell(\epsilon)$ increases monotonically with imperfection strength $\epsilon$, interpolating between a nearly ideal mode-selective mapping and a leakage-dominated regime. Here, $\mathbf{B}(\epsilon)$ denotes the leakage-mixing operator that maps the $N$ controlled input degrees of freedom into the uncontrolled subspace spanned by $\mathbf{S}$, distributing excitation among background modes populated under imperfections. Accordingly, $\mathbf{B}(\epsilon) \in \mathbb{C}^{K_s\times N}$ maps the $N$ controlled inputs into the $K_s \gg N$ uncontrolled background modes. The emergence of such speckle-like components and their statistical extreme-value behavior are well established in the theory of random interference fields [10, 7].

Within this model, the effective Gram matrix $\mathbf{G}_{\text{eff}} = \mathbf{T}^\dagger\mathbf{W}\mathbf{T}$ evolves continuously with $\epsilon$, leading to deformation, displacement, and progressive softening of stationary points in the associated phase-space landscape. Expanding Eq. (20) generates cross terms between the controlled and uncontrolled subspaces that scale as $\sqrt{\ell(1-\ell)}$. In the weak-leakage regime considered here, we assume that these cross terms either vanish approximately under

the weighting operator $\mathbf{W}$ (e.g. due to weak overlap or approximate orthogonality between controlled and uncontrolled subspaces) or average to zero under coarse graining or random-phase mixing. Under this assumption, the leading effective deformation of $\mathbf{G}_{\text{eff}}$ scales linearly with the leakage fraction $\ell$.

Rather than implying a generic loss of stationary interference solutions, increasing leakage primarily reduces basin curvature and distinctness within a given phase-space slice. The evolution of basin depths reported in Fig. 5d of the main text reflects this gradual degradation of constructive interference confined to the controlled subspace.

To examine the evolution of physical mode profiles, for each basin $B_j$ we identify the corresponding input basis vector $e_{k^*}$ at $\epsilon = 0$ by maximal overlap between the basin-locked field and the columns of $\mathbf{T}(0)$. The effective mode shapes $|\mathbf{T}(\epsilon)e_{k^*}|^2$, shown in Fig. 5e,f of the main text, then track the redistribution of energy from a localized core into a diffuse speckle-like halo as leakage increases.

These results distinguish between two complementary degradation mechanisms of phase-only control. First, leakage-induced deformation of the reduced phase-space landscape softens basins and lowers robustness to phase noise and perturbations. Second, physical mode profiles degrade through energy leakage into uncontrolled degrees of freedom. Together, these effects limit the achievable robustness and fidelity of phase-induced relocalization without requiring the disappearance of stationary interference solutions.

## Supplementary Note 6: Basin softening under leakage and asymmetry

This section describes how discrete phase-space basins emerging in a conservative, phase-only model evolve when the controlled phase subspace is coupled to uncontrolled degrees of freedom. Starting from a structured phase-space landscape that supports multiple well-defined basin centers, we introduce a phenomenological leakage parameter that reduces basin discriminability by mixing structured basin contributions with a basin-independent background. The resulting probabilistic basin description captures a continuous crossover from discrete basin identity to mixed basin membership as leakage increases. Figures S1 and S2 visualize this crossover and quantify the associated loss of basin stiffness, demonstrating that spatially localized states persist under strong leakage while their robustness and selectivity are progressively reduced.

**Phase-space landscape and conservative basins.** We consider a reduced phase-space description in terms of two relative phases $(\varphi_1, \varphi_2)$ defined on a torus. The conservative phase-space landscape shown in Fig. 3b (main draft) is constructed as a superposition of $K = 4$ localized contributions,

$$\mathcal{L}_\alpha(\varphi_1, \varphi_2) = \sum_{k=1}^{K} A_k \exp\left[-\frac{(\Delta\varphi_{1,k})^2 + (\Delta\varphi_{2,k})^2}{2\sigma_k^2}\right] + \mathcal{C}(\varphi_1, \varphi_2), \tag{21}$$

where $\Delta\varphi_{i,k}$ denotes the angular distance on the torus to the $k$th basin center, $A_k$ and $\sigma_k$ control the amplitude and width of each contribution, and $\mathcal{C}(\varphi_1, \varphi_2)$ is a weak corrugation term introducing shared structure across basins. The landscape is normalized to the interval $[0, 1]$ for visualization.

From this landscape, structured per-basin score fields $S_k^{\text{struct}}(\varphi_1, \varphi_2)$ are defined such that

$$\sum_{k=1}^{K} S_k^{\text{struct}}(\varphi_1, \varphi_2) = \mathcal{L}_\alpha(\varphi_1, \varphi_2). \tag{22}$$

In the conservative limit, discrete basins (Fig. S1b) are obtained by assigning each phase-space point to the basin with the maximum score,

$$\text{basin}(\varphi_1, \varphi_2) = \arg\max_k S_k^{\text{struct}}(\varphi_1, \varphi_2), \tag{23}$$

corresponding to robust stationary solutions accessible via phase-only control, whose robustness is governed by the local geometry of their basins of attraction, a standard concept in nonlinear dynamical systems [4, 11].

**Leakage model and softened basin scores.** To model realistic imperfections discussed in the main draft—such as unintended intermodal coupling, radiation into cladding modes, or symmetry-breaking loss—we introduce a phenomenological leakage parameter $\gamma \in [0, 1]$, distinct from the physical leakage fraction $\ell(\epsilon)$ defined in the operator model. Leakage represents coupling of the controlled phase subspace to uncontrolled degrees of freedom and is implemented by mixing the structured scores with a basin-independent background,

$$S_k(\varphi_1, \varphi_2) = (1 - \gamma)\, S_k^{\text{struct}}(\varphi_1, \varphi_2) + \gamma\, S_{\text{leak}}, \tag{24}$$

where $S_{\text{leak}}$ is a constant chosen as the spatial average of $S_k^{\text{struct}}$. Here $S_{\text{leak}}$ is defined as the phase-space average of the structural similarity $S_{\text{struct},k}(\boldsymbol{\varphi})$ over the reduced phase torus, i.e.

$$S_{\text{leak}} = \langle S_{\text{struct},k}(\boldsymbol{\varphi}) \rangle_{\boldsymbol{\varphi}} \, ,$$

where the average is taken uniformly over $\boldsymbol{\varphi}$ and, by symmetry, is independent of the output index $k$.

Importantly, this procedure preserves the underlying landscape while reducing the relative contrast between competing basins. As a result, stationary solutions persist, but basin boundaries become progressively softer. Such softening of basin boundaries under coupling to uncontrolled degrees of freedom is consistent with the general behavior of dynamical systems subject to random or weak perturbations, where stationary solutions persist while their basins of attraction become probabilistic rather than sharply defined [12].

**Probabilistic basin membership.** For finite leakage, basin identity is no longer deterministic. We therefore define probabilistic basin membership using a softmax transformation,

$$P_k(\varphi_1, \varphi_2) = \frac{\exp[S_k(\varphi_1, \varphi_2)/\tau]}{\sum_{j=1}^{K} \exp[S_j(\varphi_1, \varphi_2)/\tau]}, \tag{25}$$

which corresponds to a Gibbs-type distribution commonly used to describe probabilistic state selection in systems with competing configurations [13]. $S_k(\phi)$ denotes a basin-affinity score (e.g. proximity or local objective contribution) associated with basin $k$. Here, $\tau$ controls the sharpness of basin selection. To ensure scale-invariant behavior, $\tau$ is expressed as

$$\tau = \tau_{\text{rel}} \, \sigma_S, \tag{26}$$

with $\sigma_S$ the standard deviation of the structured scores in the conservative limit. Throughout Figs. S1 and S2, $\tau_{\text{rel}}$ is held fixed.

Figure S1c visualizes the resulting probabilistic basin membership by mapping the probabilities $P_k$ onto an RGB color mixture.

**Quantifying basin softening.** To quantify the degree of basin non-discreteness, we compute the maximum basin probability

$$P_{\max}(\varphi_1, \varphi_2) = \max_k P_k(\varphi_1, \varphi_2), \tag{27}$$

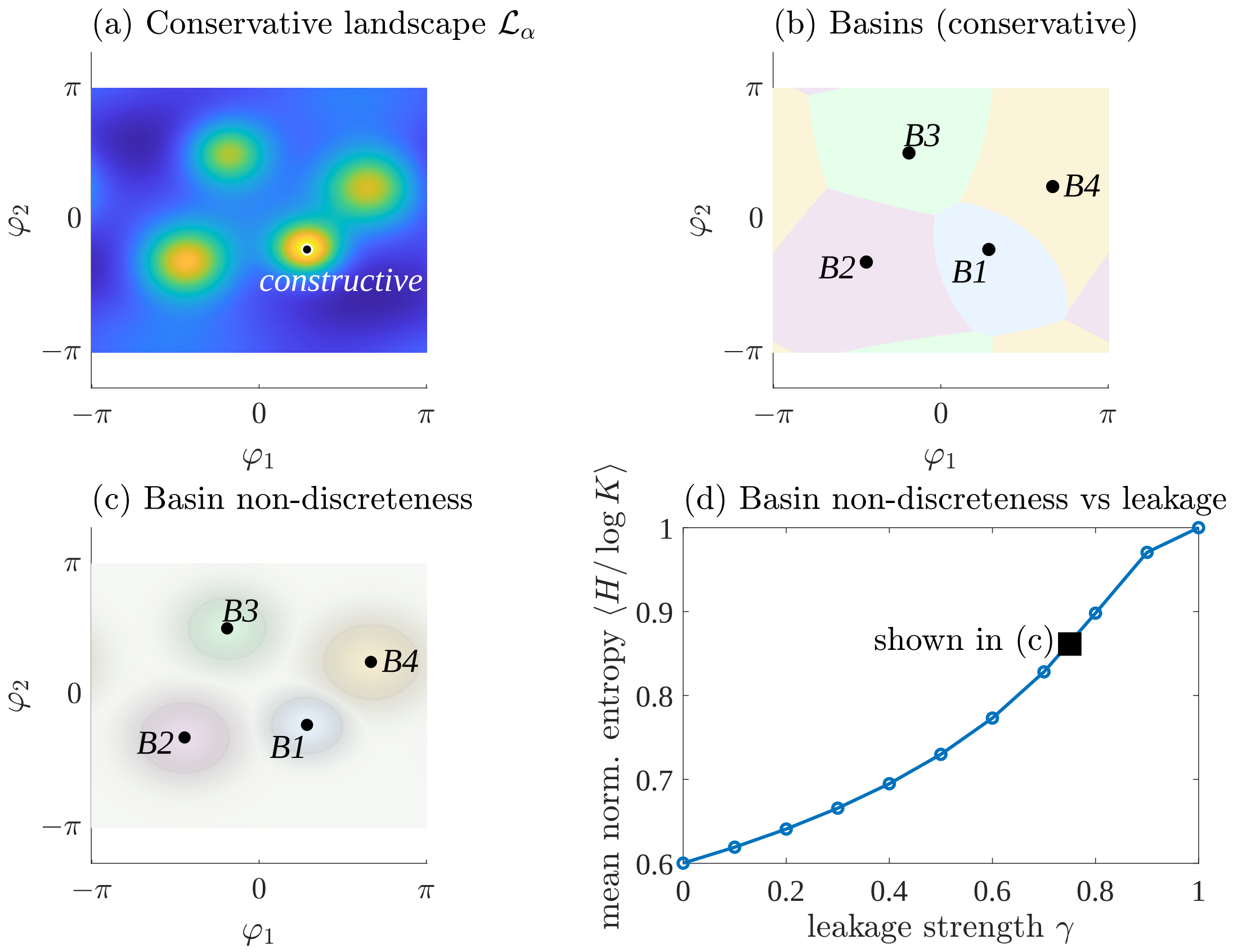


Figure S1: **Basin non-discreteness induced by leakage into uncontrolled modes.** **(a)** Conservative phase-space landscape $\mathcal{L}_\alpha(\varphi_1, \varphi_2)$ constructed from the same multimode schematic used in Fig. 3 of the main draft. **(b)** Corresponding conservative basins obtained from the argmax of the structured per-basin scores, yielding discrete basin centers $B_1$–$B_4$. **(c)** Probabilistic basin membership under finite leakage ($\gamma = 0.75$), shown as an RGB mixture weighted by basin probabilities $P_k$. Leakage is modeled as coupling of the controlled phase subspace to basin-independent background degrees of freedom. The color overlay indicates the degree of basin non-discreteness, quantified by the normalized measure defined in Eq.28, while the thin band highlights regions exceeding a fixed non-discreteness threshold. **(d)** Mean normalized entropy $\langle H/\log K\rangle$ of the basin probability distribution as a function of leakage strength $\gamma$, demonstrating a continuous crossover from discrete to mixed basin identity rather than an abrupt collapse.

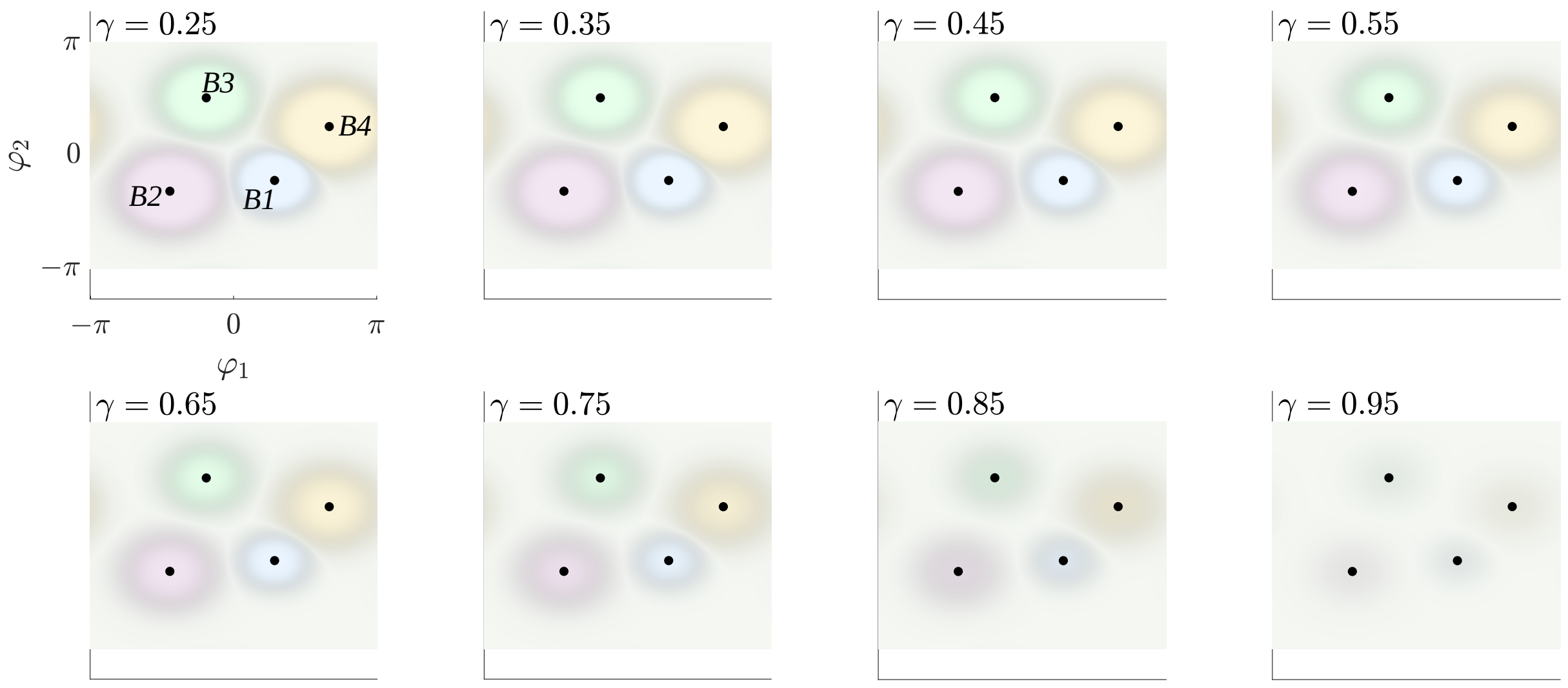


Figure S2: **Progressive softening of basin identity with increasing leakage.** Montage of panel (c) from Fig. S1 showing probabilistic basin membership for increasing leakage strength $\gamma = 0.25$–$0.95$ (left to right, top to bottom), at fixed softmax temperature $\tau_{\mathrm{rel}} = 0.35$. As leakage increases, basin boundaries progressively dissolve and basin membership becomes increasingly mixed, while residual localization persists due to underlying modal structure. The sequence illustrates that strong leakage suppresses basin discriminability continuously, rather than eliminating spatial localization or confinement.

and define a normalized non-discreteness measure,

$$\mathrm{nonDisc}(\varphi_1, \varphi_2) = 1 - \frac{P_{\max} - 1/K}{1 - 1/K}, \tag{28}$$

which ranges from 0 (fully discrete basin identity) to 1 (uniform basin mixing). This quantity is overlaid in Fig. S1c to highlight regions where basin competition is strong.

In addition, we compute the Shannon entropy of the basin probability distribution,

$$H(\varphi_1, \varphi_2) = -\sum_{k=1}^{K} P_k \log P_k, \tag{29}$$

normalized by $\log K$. Figure S1d shows the mean normalized entropy $\langle H/\log K\rangle$ as a function of leakage strength $\gamma$, demonstrating a smooth crossover from discrete to mixed basin identity.

**Progressive basin softening.** Figure S2 presents a montage of Fig.S1c for increasing leakage strength $\gamma = 0.25$–$0.95$. As leakage increases, basin boundaries dissolve and basin

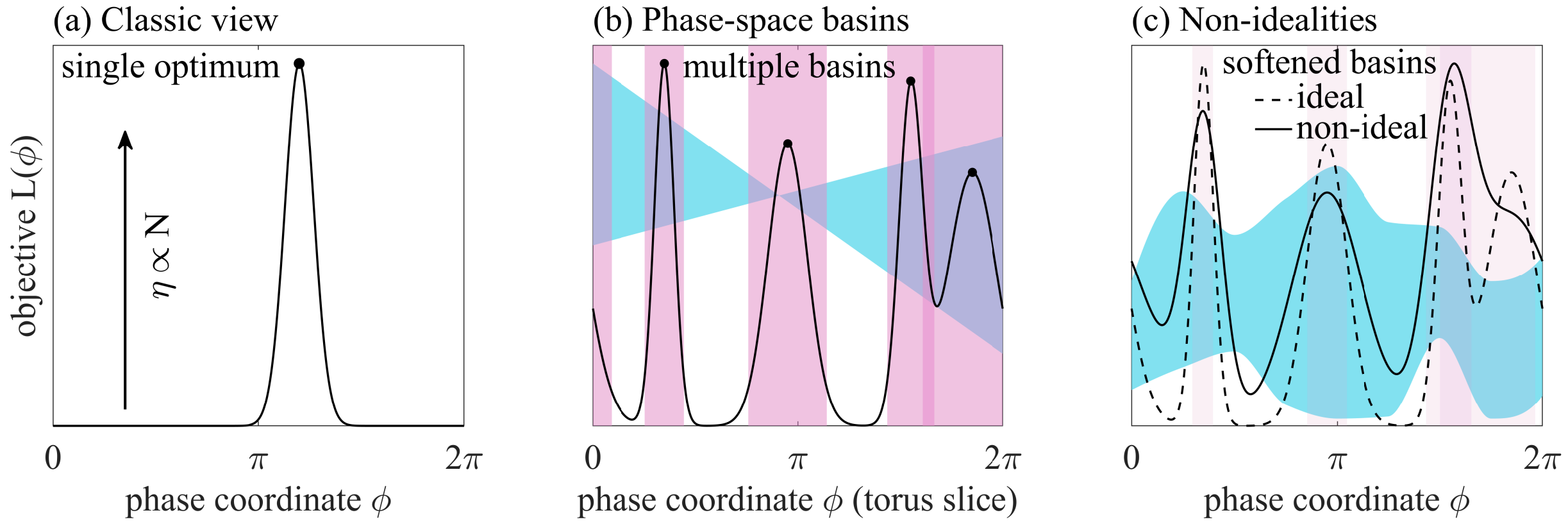


Figure S3: **Bounds versus basin geometry in phase space. (a)** Classical wavefront-shaping picture: a single optimum whose peak height is constrained by the enhancement bound $\eta \propto N$. **(b)** Phase-space view on a torus slice showing multiple optima with comparable peak heights but different basin extents (vertical shaded bands). The slanted background indicates a slowly varying interference contribution from uncontrolled modal subspaces, which modulates local curvature without introducing additional stationary points. **(c)** Non-ideal conditions: peak heights remain comparable to the ideal case (dashed), while basin extents shrink and soften (solid), reflecting reduced robustness and likelihood of convergence. A more irregular background illustrates increased leakage-induced interference. Overall, bounds constrain achievable peak height, whereas basin geometry governs robustness and optimization probability.

membership becomes increasingly probabilistic. Nevertheless, residual spatial localization persists even at strong leakage, reflecting the continued existence of stationary solutions supported generically by phase-space topology. These figures therefore directly illustrate that performance degradation arises from basin softening rather than from the disappearance of stationary solutions.

**Comparison of classical and phase-space basin geometries.** Figure S3 provides a schematic comparison between conventional enhancement-bound arguments and a phase-space interpretation based on basin geometry. In the classical picture, illustrated in Fig. S3a, wavefront shaping is characterized by a single dominant optimum whose peak height scales with the number of controlled degrees of freedom, $\eta \propto N$. Such bounds quantify the maximum achievable enhancement but do not address how likely this optimum is to be reached under realistic optimization dynamics.

Figure S3b reframes the problem in phase space. On a one-dimensional torus slice of the control manifold, multiple optima coexist with comparable peak heights, yet their associated basins of attraction differ substantially in extent. These basin sizes are represented by vertical

shaded bands, emphasizing that optimization dynamics depend not only on peak height but also on the volume of phase space that funnels trajectories toward each optimum. The slanted background shading represents a slowly varying interference contribution from uncontrolled modal subspaces. This contribution modifies local curvature and basin geometry without introducing additional stationary points, and therefore does not affect enhancement bounds.

Figure S3c illustrates the impact of non-idealities such as partial mode control, noise, or leakage into uncontrolled channels. While the locations and approximate heights of the optima persist (dashed versus solid curves), basin extents shrink and soften, indicating reduced robustness to perturbations and sensitivity to initialization. The more irregular background shading reflects stronger, spatially structured interference from uncontrolled subspaces, further distorting basin geometry while leaving peak heights largely unchanged.

Taken together, Fig. S3 highlights a key distinction: enhancement bounds constrain what peak heights are achievable, whereas basin geometry determines robustness, convergence probability, and experimental reproducibility. This separation explains why systems with similar enhancement limits can exhibit markedly different optimization performance under non-ideal conditions.

## Supplementary Note 7: Extended geometric consequences of compact phase-only control

The scaling law stated in the main text represents the simplest observable manifestation of compact phase-only control under leakage into uncontrolled degrees of freedom. Here we formulate a more general geometric deformation principle that governs not only peak contrast, but also basin robustness and basin identity. These quantities are structurally coupled by the same dimensional mismatch between the controlled toroidal manifold and the leakage-populated uncontrolled subspace.

**Compact-control deformation principle.** Let the localization functional be defined on the reduced (gauge-fixed) phase manifold[3, 6, 7]

$$\mathcal{L} : \mathbb{T}^{N-1} \to \mathbb{R}. \tag{30}$$

Under phase-only actuation, $\mathcal{L}$ is continuous on the compact torus $\mathbb{T}^{N-1}$ and therefore admits isolated stationary points corresponding to localized interference configurations.

In the operator formulation, quadratic observables take the form

$$\mathcal{L}(\phi) = x^\dagger(\phi)\, G_{\text{eff}}\, x(\phi), \tag{31}$$

where $x(\phi) \in \mathbb{C}^N$ denotes the normalized phase-only input state with fixed amplitudes and tunable relative phases, e.g. $x_j(\phi) = |c_j| e^{i\tilde{\phi}_j}/\|c\|$. The effective Gram operator

$$G_{\text{eff}} = T^\dagger W T \tag{32}$$

incorporates the system transmission operator $T$ and the Hermitian weighting operator $W$ defining the measured quadratic observable (e.g. a spatial window or projection).

We decompose

$$G_{\text{eff}} = G_c + \delta G, \tag{33}$$

where $G_c$ acts within the controlled subspace and $\delta G$ encodes mixing with uncontrolled degrees of freedom. For weak leakage fraction $\ell$, $\|\delta G\|$ scales linearly with $\ell$.

Because peak contrast, basin curvature, and basin identity all derive from the same operator $G_{\text{eff}}$, they undergo a jointly constrained deformation under $\delta G$. The following corollaries make this structure explicit.

### A. Contrast–robustness tradeoff

Let $\phi^\star$ denote a stable stationary point of $\mathcal{L}$. Expanding to second order,

$$\mathcal{L}(\phi^\star + \delta\phi) \simeq \mathcal{L}(\phi^\star) - \frac{1}{2}\, \delta\phi^{\mathrm{T}} K\, \delta\phi, \tag{34}$$

where $K = -H$ is the basin stiffness matrix and $H$ the Hessian evaluated at $\phi^\star$.

Under perturbation $\delta G$, both the maximal contrast

$$\eta_{\max} = \frac{\mathcal{L}(\phi^\star)}{\mathcal{L}_{\text{ideal}}}, \tag{35}$$

where $\mathcal{L}_{\text{ideal}}$ denotes the optimal value in the leakage-free limit, and the stiffness spectrum $\{\kappa_i\}$ of $K$ deform at leading order in $\ell(N-1)$. In particular,

$$\eta_{\max} \lesssim 1 - C\, \ell(N-1), \tag{36}$$

while the trace of the stiffness matrix satisfies

$$\mathrm{Tr}(K) = \mathrm{Tr}(K_c) + \mathcal{O}(\ell(N-1)) \,, \tag{37}$$

with the same geometric overlap parameter $C$ governing both corrections. Here $C$ depends only on the conditioning and overlap between the controlled image of $T$ and the leakage-populated uncontrolled subspace (see Supplementary Note 9).

Because the local quadratic tolerance volume scales inversely with curvature,

$$\mathcal{V}_\delta(\phi^\star) \propto \frac{1}{\sqrt{\det K}}, \tag{38}$$

curvature softening alone enlarges the local level-set neighborhood associated with a fixed tolerated localization loss $\delta$. However, leakage simultaneously reduces peak contrast and basin discriminability, so the experimentally useful robustness region is not determined by curvature alone, but by the combined requirements of localization fidelity, basin separability, and tolerance to perturbations. Peak localization and experimentally accessible robustness therefore cannot be independently optimized under phase-only control, as both are constrained by the same dimensional mismatch parameter.

### B. Loss of basin discreteness and entropy crossover

In the conservative limit, each phase-space point belongs deterministically to a single basin via an argmax rule. Under leakage-induced mixing, structured per-basin scores are perturbed and basin identity becomes probabilistic.

Define basin probabilities via a softmax transformation,

$$P_k(\phi) = \frac{\exp[S_k(\phi)/\tau]}{\sum_j \exp[S_j(\phi)/\tau]}, \tag{39}$$

and the normalized basin entropy

$$\bar{H} = \frac{\left\langle -\sum_{k=1}^{K} P_k(\varphi) \log P_k(\varphi) \right\rangle_\varphi}{\log K}. \tag{40}$$

In the absence of leakage, $\bar{H} \to 0$, reflecting discrete basin identity. For finite leakage,

$$\bar{H} = f(\ell(N-1)) \,, \tag{41}$$

with $f$ monotone increasing to leading order. Thus leakage induces a continuous crossover from deterministic basin identity to mixed basin membership without eliminating stationary interference solutions.

## Supplementary Note 8: Experimental measurement of basin robustness and curvature under phase-only control

This note provides a minimal, platform-independent experimental protocol for extracting basin stiffness and curvature under phase-only control. The protocol requires only phase modulation and intensity measurements and does not rely on transmission-matrix reconstruction, full-field access, or iterative optimization.

**Experimental identification of a basin center.** In an experiment, a basin center corresponds to the phase configuration that maximizes a chosen localization metric, such as the intensity at a target position or the power coupled into a localized output mode. In practice, this configuration is obtained by constructive phase alignment, phase conjugation, or direct maximization of the measured intensity. This maximum defines the basin center $\boldsymbol{\phi}^\star$, about which local robustness and curvature are subsequently characterized.

**Measurement of basin robustness from phase perturbations.** In the vicinity of a stable basin center, the localization functional varies smoothly with small phase perturbations and admits a quadratic expansion,

$$L(\boldsymbol{\phi}^\star + \delta\boldsymbol{\phi}) \simeq L(\boldsymbol{\phi}^\star) - \frac{1}{2}\,\delta\boldsymbol{\phi}^\mathsf{T} K\,\delta\boldsymbol{\phi}, \tag{42}$$

where $K \equiv -H$ is the basin stiffness matrix and $H$ is the Hessian of $L$ evaluated at $\boldsymbol{\phi}^\star$. The matrix $K$ encodes the local curvature of the phase-space landscape and determines the sensitivity of the localized state to phase perturbations.

Experimentally, basin robustness is accessed by applying small random phase perturbations $\delta\boldsymbol{\phi}$ with known covariance $\Sigma = \langle \delta\boldsymbol{\phi}\,\delta\boldsymbol{\phi}^\mathsf{T}\rangle$ around $\boldsymbol{\phi}^\star$ and measuring the resulting change in the localization functional. Averaging over multiple realizations yields

$$\langle \Delta L\rangle \equiv \langle L(\boldsymbol{\phi}^\star + \delta\boldsymbol{\phi}) - L(\boldsymbol{\phi}^\star)\rangle \simeq -\frac{1}{2}\,\mathrm{Tr}(K\Sigma). \tag{43}$$

For fixed and known perturbation statistics, the measured quantity $\langle \Delta L\rangle$ provides direct

access to the noise-weighted basin stiffness $\mathrm{Tr}(K\Sigma)$. For isotropic perturbations, $\Sigma = \sigma^2 I$, this reduces to a measurement of the scalar stiffness $\mathrm{Tr}(K)$ up to the known factor $\sigma^2$.

**Broader applicability.** The phase-space perspective developed here establishes basin geometry as a compact and experimentally accessible descriptor of multimode interference under constrained control. While the framework provides direct access only to local geometric properties of the phase-space landscape, more global system characteristics—such as asymmetry and leakage into uncontrolled degrees of freedom—are not measured directly but leave systematic signatures in the deformation, softening, and variability of basin structure across localized states and system parameters. This separation between locally accessible observables and globally inferred properties clarifies both the capabilities and the limits of phase-only control, and suggests a route toward comparative characterization of complex multimode systems through geometric diagnostics, even in regimes where full-field control or operator reconstruction is impractical.

## Supplementary Note 9: Cross-platform realizations of leakage and geometric scaling

The geometric scaling law states that for a coherent system with phase-only actuation on $N$ modes and leakage fraction $\ell$ into uncontrolled degrees of freedom, the maximal achievable localization or contrast obeys, to leading order,

$$\eta_{\max} \lesssim 1 - C\,\ell(N-1), \tag{44}$$

where $C$ depends only on the geometric overlap and conditioning between the controlled image of the system operator and the uncontrolled leakage subspace.

The perturbative regime is governed by the dimensionless parameter

$$\Xi \equiv \ell(N-1), \tag{45}$$

which quantifies the dimensional mismatch between compact phase-only control and leakage into uncontrolled degrees of freedom. For $\Xi \ll 1$, contrast reduction, basin curvature softening, and entropy growth remain linear in $\Xi$. When $\Xi \sim O(1)$, a crossover occurs to a leakage-dominated regime in which stationary interference configurations persist by compactness of the $(N-1)$-dimensional phase manifold, but their attainable contrast and

robustness undergo $O(1)$ reductions independent of further increases in $N$. For larger leakage fractions, higher-order corrections in $\ell$ become non-negligible and can modify the linear scaling, although stationary configurations persist by compactness.

Importantly, this result does not rely on a specific optical architecture. It applies to any system in which: (i) control is restricted to relative phases on a compact manifold $U(1)^{N-1}$ (or an equivalent toroidal phase space after gauge reduction), and (ii) the observable of interest is a Hermitian quadratic functional of the coherent field or state.

If leakage coupling is isotropic (e.g. Haar-random with respect to the controlled image), the geometric coefficient $C$ concentrates around its mean with fluctuations that decrease with $N$. In this case, the linear dependence on $(N-1)$ arises generically from dimensional mismatch rather than device-specific structure, defining a universality class for compact phase-only control under random leakage.

This note clarifies (i) how the leakage fraction $\ell$ is defined in a platform-independent operator framework, (ii) how the coefficient $C$ arises geometrically, and (iii) how the bound specializes across representative platforms spanning classical waves, many-body systems, and quantum coherent devices.

**Projection-operator formulation of leakage.** Let the physical system be described by the linear transmission operator

$$T : \mathbb{C}^N \to \mathbb{C}^M, \tag{46}$$

which maps controlled input coefficients to output field amplitudes. This is the same operator appearing in the Gram construction $G = T^\dagger W T$, acting on phase-only controlled inputs $x(\varphi)$ with fixed amplitudes and variable relative phases. The output field or state is

$$\mathbf{E} = T\mathbf{x}. \tag{47}$$

The accessible control manifold is the compact torus

$$\mathcal{M} \cong U(1)^{N-1}, \tag{48}$$

arising from the gauge redundancy of global phase.

We decompose the output Hilbert space as

$$\mathcal{H} = \mathcal{H}_c \oplus \mathcal{H}_u, \tag{49}$$

with orthogonal projectors $P_c$ and $P_u$.

The leakage fraction is defined operationally as

$$\ell(\boldsymbol{\phi}) \equiv \frac{\|P_u T\mathbf{x}(\boldsymbol{\phi})\|^2}{\|T\mathbf{x}(\boldsymbol{\phi})\|^2}. \tag{50}$$

We distinguish the imperfection-indexed leakage $\ell(\epsilon)$ (Note 5), the phase-dependent leakage $\ell(\phi)$, and the phenomenological mixing parameter $\gamma$ (Note 6), which quantify related but distinct aspects of leakage-induced mixing. Under ideal constructive phase alignment, $P_u T\mathbf{x} = 0$. Imperfections—such as disorder, dissipation, decoherence, uncontrolled coupling, or dimensional mismatch—generate a nonzero uncontrolled component. For weak leakage, observables may be expanded perturbatively in $\ell$.

**Origin of the scaling coefficient $C$.** Let the performance metric be a quadratic observable,

$$\mathcal{L} = \mathbf{x}^\dagger G_{\text{eff}}\mathbf{x}, \qquad G_{\text{eff}} = T^\dagger W T, \tag{51}$$

where $W$ is a Hermitian weighting operator defining the measured observable (e.g., intensity in a region, projection onto a mode, spin projection, current operator, or interference visibility).

Under weak leakage,

$$G_{\text{eff}} = G_c + \delta G, \tag{52}$$

where $G_c$ acts within the controlled subspace and $\delta G$ encodes mixing with uncontrolled degrees of freedom.

Let $\lambda_{\max}$ denote the maximal eigenvalue of the effective Gram operator $G_{\text{eff}}$, which determines the optimal quadratic objective under unconstrained complex control. For weak leakage, the deformation of the optimal objective is governed by the leakage-induced perturbation $\delta G$ of the effective Gram operator. Standard eigenvalue perturbation theory implies the bound

$$|\delta\lambda_{\max}| \le \|\delta G\|, \tag{53}$$

where $\|\cdot\|$ denotes the operator norm. In the weak-leakage regime considered here, and under the assumption that controlled–uncontrolled cross terms either vanish approximately under the weighting operator or average out under coarse graining, the leading perturbation scales linearly with the leakage fraction $\ell$.

When leakage is distributed across the $(N-1)$ independent controlled phase directions, the

typical degradation of the optimal quadratic objective therefore scales as

$$|\delta\lambda_{\max}| \sim C\,\ell\,(N-1), \tag{54}$$

which yields the linear dependence in Eq. (44).

The coefficient $C$ depends only on the geometric overlap structure between $T(\mathcal{H}_c)$ and $\mathcal{H}_u$, i.e. on the conditioning of controlled-to-uncontrolled coupling. It is independent of device architecture, inversion protocol, or optimization strategy.

The scaling therefore reflects a structural incompatibility between:

1. Compact phase-only control on $U(1)^{N-1}$, and
2. Leakage into an uncontrolled Hilbert subspace whose effective dimension grows with $N$.

**Universality beyond optical wave systems.** The operator formulation above applies not only to classical wave propagation but to any coherent system where amplitudes are fixed and relative phases constitute the only accessible controls. Examples include spin ensembles with RF phase control in nuclear magnetic resonance and quantum control landscapes [14], Josephson junction arrays with tunable superconducting phase differences [15], Bose–Einstein condensates in optical lattices with controlled inter-site phase coherence [16], and coupled nonlinear oscillator networks exhibiting phase synchronization phenomena [17].

In all such cases:

- The accessible control manifold reduces (after gauge fixing) to a compact torus.
- at least extrema persist by compactness; non-degenerate stationary points deform continuously under weak perturbations
- Leakage into uncontrolled degrees of freedom limits contrast through dimensional mismatch.

Thus, Eq. (44) expresses a geometric constraint on phase-restricted coherent control, not a property specific to optical interference.

**Representative platform instantiations.** Although the form of Eq. (44) is universal, the physical origin of $\ell$ and the conditioning factor $C$ vary across systems. Table S1 summarizes representative cases spanning classical, many-body, and quantum coherent platforms.

Table S1: Cross-platform realization of leakage fraction $\ell$ and geometric scaling.

| **Platform** | **Controlled DOF** | **Source of Leakage $\ell$** | **Interpretation of scaling** |
|---|---|---|---|
| Multimode fiber[18] | Input modal phases | Polarization mixing, modal dispersion | Spectral and modal mismatch limits relocalization |
| Integrated mesh/MMI[19] | Phase shifters | Fabrication disorder, non-unitary loss | Disorder-induced subspace mixing |
| Disordered scattering medium[8] | SLM phase pixels | Uncontrolled amplitude DOF, temporal decorrelation | Contrast limited by field correlation |
| Microwave[20] | Phased ports | Dissipation, radiation loss | Loss-to-coupling ratio limits contrast |
| Acoustic cavity [21] | Phased transducer array | Dissipation, radiation loss, environmental coupling | Loss-to-coupling ratio and modal overlap limit focal contrast |
| Spin ensembles (NMR/ESR)[22] | Relative spin phases | Dephasing, bath coupling | Loss of collective spin contrast |
| Superconducting circuits[15] | Josephson phase differences | Quasiparticle loss, parasitic modes | Reduced interference of collective current states |
| Cold-atom phase arrays[16] | Condensate phase differences | Thermal fraction, higher-band excitation | Reduced interference visibility |
| Coupled oscillator networks[17] | Oscillator phases | Noise, non-participating modes | Reduced synchronization contrast |
| Quantum interferometers[14] | Path phase shifts | Decoherence channels | Fringe visibility suppression |

In each case, $\ell$ quantifies coherent power redistributed into degrees of freedom not correctable by phase-only actuation. The scaling coefficient $C$ reflects geometric overlap between controlled and uncontrolled subspaces.

**Beyond "dimensional dilution".** The linear dependence on $\ell(N-1)$ is not a trivial energy-sharing effect. Its nontrivial content lies in the coexistence of two structural facts:

1. Compactness of the $(N-1)$-torus generically enforces a finite set of isolated stationary interference configurations, independent of leakage.

2. Dimensional mismatch between compact phase control and leakage-populated uncontrolled subspaces intrinsically limits achievable contrast of those stationary configurations.

Increasing $N$ therefore has a dual geometric consequence:

- It increases multiplicity of stationary basins (via compactness).
- It amplifies sensitivity to leakage (via dimensional mismatch).

The scaling constraint thus emerges from geometric incompatibility between compact phase-only control and expanding uncontrolled dimensionality. It is a structural property of coherent systems with restricted phase actuation, not a device-specific limitation.

**Parametric divergence from classical enhancement scaling.** Classical phase-only enhancement bounds—derived for optimization over the full complex coefficient space of the controlled modes and typically assuming access to the effective transmission operator—predict a contrast that scales proportionally with the number of controlled degrees of freedom $N$ under idealized full control. In contrast, the compact-control bound derived here yields $\eta_{\max} \lesssim 1 - C\,\ell(N-1)$ when modal amplitudes are fixed and actuation is restricted strictly to relative phases. For any leakage fraction $\ell$ that remains independent of $N$, the two predictions therefore diverge parametrically as $N$ increases: enhancement theory predicts monotonic improvement with $N$, whereas compact phase-manifold geometry predicts a leakage-dominated regime once $\ell(N-1) \gtrsim O(1)$. In this regime, higher-order corrections in $\ell$ become non-negligible and contrast reduction is no longer perturbative in $\ell$; nevertheless, stationary interference configurations persist by compactness while their contrast, stiffness spectrum, and basin identity undergo order-unity degradation. The divergence reflects dimensional mismatch between the $(N-1)$-dimensional toroidal control manifold and leakage-populated uncontrolled subspaces, rather than limitations of a particular optimization protocol.

## Supplementary Note 10: Comparison with existing multimode control and characterization frameworks

Having established in Supplementary Note 9 that the geometric scaling constraint arises generically from compact phase-only control on a $U(1)^{N-1}$ manifold—independently of spe-

cific physical platform—we now position the phase-space basin framework relative to existing approaches to coherent multimode control.

Multiple theoretical frameworks exist for understanding and controlling coherent multimode wave systems, including control-landscape theory, wavefront shaping, transmission-matrix methods, and eigenchannel-based approaches. These frameworks address different questions and assume different levels of control access and system knowledge. The present phase-space basin framework is complementary: rather than synthesizing a desired field or identifying globally optimal channels, it characterizes the existence, multiplicity, and robustness of stationary interference states that are intrinsically accessible under static phase-only control. In this regime, compactness of the control manifold enforces discrete solutions, and robustness and scaling are governed by geometric basin properties rather than by optimization performance, calibration fidelity, or algorithmic convergence. Table S2 summarizes these distinctions.

Table S2: Comparison of frameworks for coherent multimode control

| **Framework** | **Primary question addressed** | **Control degrees of freedom** | **Requires explicit system knowledge?** | **Nature of solutions** | **Meaning of robustness** | **Scaling insight** |
|---|---|---|---|---|---|---|
| Control-landscape theory[23, 22] | Is a trap-free global optimum reachable under admissible controls? | High-dimensional, often unconstrained; time-dependent | Often yes (model or Hamiltonian knowledge) | Global extrema of a cost functional | Algorithmic stability and convergence under optimization | Generally not central; depends on controllability assumptions |
| Wavefront shaping[24, 8] | How much enhancement or focusing can be achieved at a target? | Phase-only (typically), or amplitude+phase control | Often requires feedback or transmission-matrix measurement | Optimized focal patterns or enhancements | Sensitivity to noise, feedback precision, or modulation resolution | Asymptotic enhancement bounds with system size |
| Transmission-matrix methods[25, 26] | Can the medium be inverted or precompensated? | Field control (phase or complex) | Yes (transmission/scattering matrix measurement) | Deterministic refocusing or imaging states | Calibration fidelity and matrix conditioning | Condition number and rank of the transmission matrix |
| Eigenchannels/ principal modes[9, 27] | What special invariant or optimal channels exist? | Modal superpositions | Yes (scattering or propagation operator) | Eigenstates with extremal or invariant properties | Disorder robustness of specific channels | Statistical scaling of channel properties |
| **Phase-space-basin framework (this work)** | **Which stationary interference states exist and how robust are they?** | **Static phase-only control on a compact manifold** | **No** | **Discrete stationary states with finite phase-space basins** | **Geometric basin stiffness and curvature** | **Scaling governed by basin geometry; limits set by softening/leakage** |

## Supplementary Note 11: Quantum realization of compact phase-manifold scaling

The geometric scaling constraint derived in the main text applies to any coherent system in which control is restricted to relative phases and observables are Hermitian quadratic forms. Finite-dimensional quantum coherent systems under phase-only actuation provide a direct realization of this structure within the framework of quantum control landscapes [14].

**Phase-only control of a finite-dimensional quantum state.** Consider a normalized pure state in an $N$-dimensional Hilbert space,

$$|\psi(\boldsymbol{\phi})\rangle = \sum_{m=1}^{N} c_m e^{i\phi_m} |m\rangle, \tag{55}$$

where $\{|m\rangle\}$ is a fixed orthonormal basis, $c_m \in \mathbb{C}$ are fixed amplitudes encoding state preparation or system coupling, and $\phi_m$ are externally controllable phases.

Removing the physically irrelevant global phase reduces the independent control parameters to an $(N-1)$-dimensional torus,

$$\mathcal{M}_\phi \cong U(1)^{N-1},$$

identical to the compact phase manifold discussed in Section 2 of the main text. Each point on $\mathcal{M}_\phi$ corresponds to a physically distinct pure state modulo global phase.

**Hermitian quadratic observables and phase-space landscapes.** Let $\hat{O}$ be a Hermitian operator representing a measurable observable (e.g., a projector, spin component, current operator, or interference observable). The expectation value

$$\mathcal{L}(\boldsymbol{\phi}) = \langle\psi(\boldsymbol{\phi})|\hat{O}|\psi(\boldsymbol{\phi})\rangle \tag{56}$$

defines a real-valued functional over the compact phase manifold.

Substituting Eq. (55) into Eq. (56) yields

$$\mathcal{L}(\boldsymbol{\phi}) = \mathbf{x}^\dagger(\boldsymbol{\phi})\, G_{\text{eff}}\, \mathbf{x}(\boldsymbol{\phi}), \tag{57}$$

where

$$\mathbf{x}(\boldsymbol{\phi}) = (c_1 e^{i\phi_1}, \ldots, c_N e^{i\phi_N})^T,$$

and

$$G_{\text{eff}} = \hat{O} \tag{58}$$

expressed in the chosen basis.

Equation (57) is formally identical to the quadratic localization functional used in the classical wave analysis. The expectation value therefore defines a smooth phase-space landscape over the compact manifold $U(1)^{N-1}$.

By continuity on a compact manifold, $\mathcal{L}(\boldsymbol{\phi})$ generically admits a finite set of isolated stationary points, corresponding to phase configurations that extremize the observable expectation value. These stationary points define discrete basins of attraction on the compact phase manifold, directly analogous to the classical interference basins described in the main text.

**Curvature, robustness, and basin stiffness.** Near a stationary configuration $\boldsymbol{\phi}^{\star}$, the observable admits a quadratic expansion

$$\mathcal{L}(\boldsymbol{\phi}^{\star} + \delta\boldsymbol{\phi}) \approx \mathcal{L}^{\star} + \frac{1}{2}\delta\boldsymbol{\phi}^{T} H \, \delta\boldsymbol{\phi}, \tag{59}$$

where $H$ is the Hessian matrix with respect to reduced phase coordinates.

The eigenvalues of $-H$ define a stiffness spectrum characterizing sensitivity of the observable to phase perturbations. As in the classical case, basin curvature quantifies robustness of the extremal quantum state to phase noise, control imprecision, or weak coupling perturbations. Importantly, this curvature depends only on the effective quadratic form and not on the optimization protocol used to reach the stationary configuration.

**Leakage and dimensional mismatch.** In realistic quantum systems, the controlled $N$-dimensional subspace may be weakly coupled to additional uncontrolled degrees of freedom, such as environmental modes or spectator states. Weak system–environment coupling modifies the effective observable structure and induces reduced dynamics within the controlled subspace [28]. Let the full Hilbert space decompose as

$$\mathcal{H} = \mathcal{H}_c \oplus \mathcal{H}_u,$$

with projectors $P_c$ and $P_u$ onto controlled and uncontrolled subspaces.

Leakage into $\mathcal{H}_u$ modifies the effective observable in the controlled subspace through

$$G_{\text{eff}} = P_c \hat{O} P_c + \delta G, \tag{60}$$

where $\delta G$ encodes coupling-induced mixing with uncontrolled states.

As shown in the classical derivation, such leakage acts as a smooth deformation of the quadratic form on the compact phase manifold. While stationary phase configurations persist by compactness, their achievable expectation-value contrast is intrinsically limited by dimensional mismatch between the $(N-1)$-dimensional control manifold and leakage-populated uncontrolled subspaces.

To leading order in weak leakage fraction $\ell$, the maximal achievable contrast obeys the same geometric scaling structure derived in the main text,

$$\mathcal{L}_{\max} \lesssim \mathcal{L}_0 - C\,\ell(N-1), \tag{61}$$

where $C$ depends only on overlap structure between controlled and leakage subspaces.

**Implications.** This construction demonstrates that compact phase-space basin geometry is not restricted to classical wave interference. Any finite-dimensional quantum coherent system with phase-only control and Hermitian quadratic observables exhibits:

1. Discrete stationary interference configurations enforced by compactness of $U(1)^{N-1}$,
2. Basin-defined robustness governed by local phase-space curvature,
3. Intrinsic scaling limits arising from dimensional mismatch with leakage-populated subspaces.

The geometric scaling constraint therefore reflects a structural incompatibility between compact phase-only control and increasing uncontrolled dimensionality, independent of whether the underlying system is classical or quantum.

## Supplementary Note 12: Experimental protocol for phase-space sampling and basin identification

The experimental identification of phase-space basins is performed using a voltage-driven three-mode photonic lantern system [29], in which relative modal phases are controlled via piezoelectric actuation, as shown in the schematic of Figure S4. One global phase is irrelevant, leaving two independent relative phases, implemented via control voltages $(V_y, V_z)$ with $V_x$ fixed. This defines a two-dimensional experimentally accessible phase space.

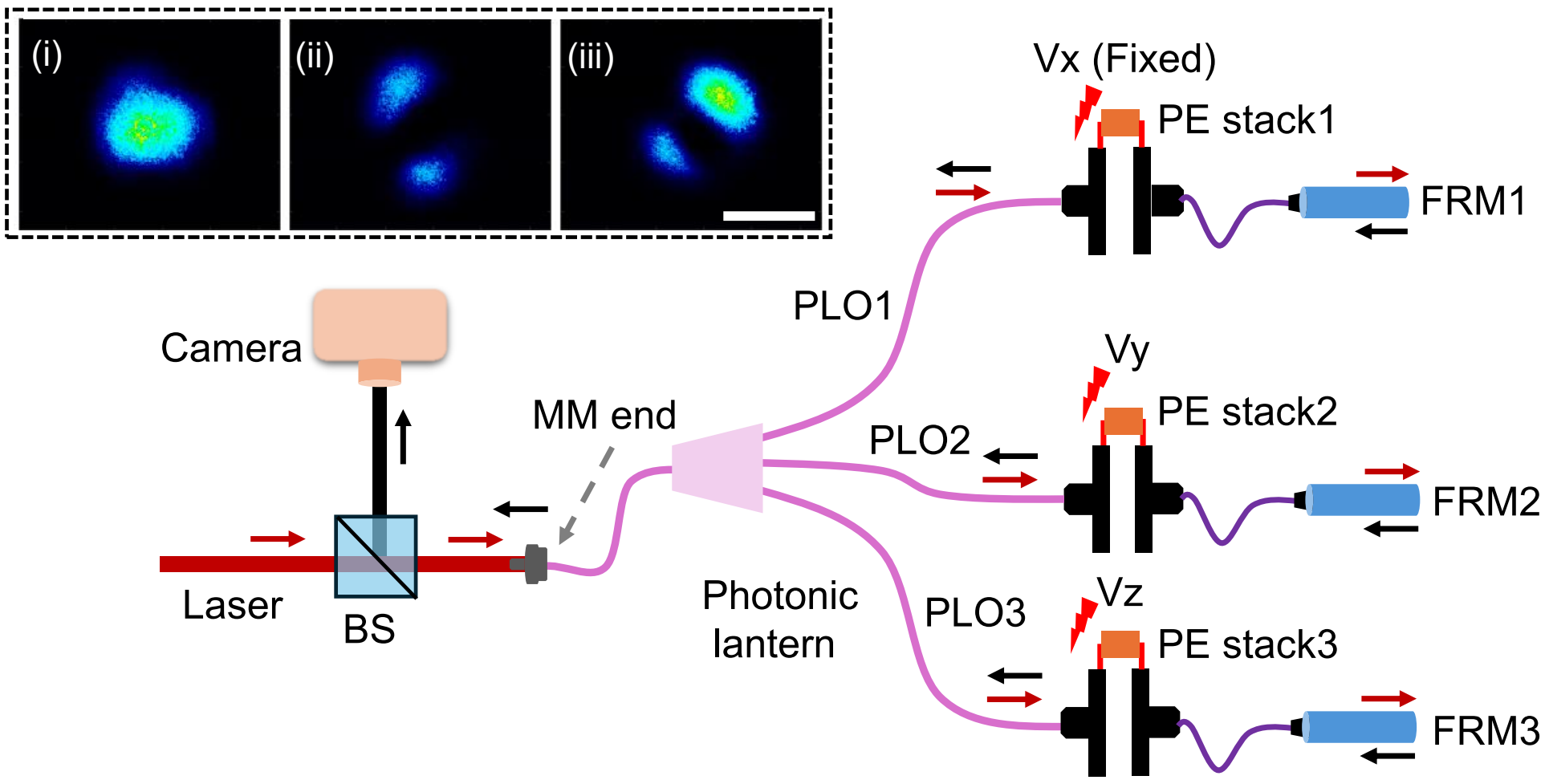


Figure S4: **Experimental implementation of phase-only control in the photonic lantern platform using coarse-voltage scan.** A coherent laser beam is coupled into a three-mode photonic lantern through a beam splitter (BS), with the transmitted light launched into the multimode end of the lantern. At each photonic lantern output (PLO), an optical delay line and piezoelectric (PE) actuator are incorporated. Phase-only control is implemented by applying voltages $(V_x, V_y, V_z)$, which act as proxies for relative phase modulation; during the coarse phase scans, $V_x$ is held fixed while $(V_y, V_z)$ are varied to sample the experimentally accessible phase space. Faraday reflecting mirrors (FRMs) return the optical fields through the same paths, enabling coherent recombination within the lantern, while the resulting mode profiles are recorded by a camera positioned at the reflection arm of the BS. Red arrows indicate forward propagation, while black arrows indicate the reflected return paths used for interference-based localization measurements. Inset- modes ($LP_{01}$ (i), $LP_{11a}$ (ii), and $LP_{11b}$ (iii)) supported by multimode end of the lantern. Scale bar, 0.5 mm.

Rather than employing iterative optimization, the phase space is sampled through continuous on-the-fly voltage scans. During these scans, output intensity distributions are recorded frame-by-frame, with each frame corresponding to a physically realizable operating point. Representative sampled frames and their corresponding energy traces are shown in Figure S5, illustrating the evolution between localized and intermediate interference states during the scan. This approach enables direct experimental sampling of the accessible phase-space landscape through voltage-controlled phase proxies, allowing identification of distinct localization basins and visualization of basin-boundary transitions through abrupt changes in the measured output patterns.

A key aspect of the protocol is the temporal sampling strategy. The output is recorded at 5 frames per second, while control voltages are updated every 2 s. This ensures that each voltage configuration is sampled over approximately 10 consecutive frames, so that each point in phase space is represented by a short temporal sequence rather than a single

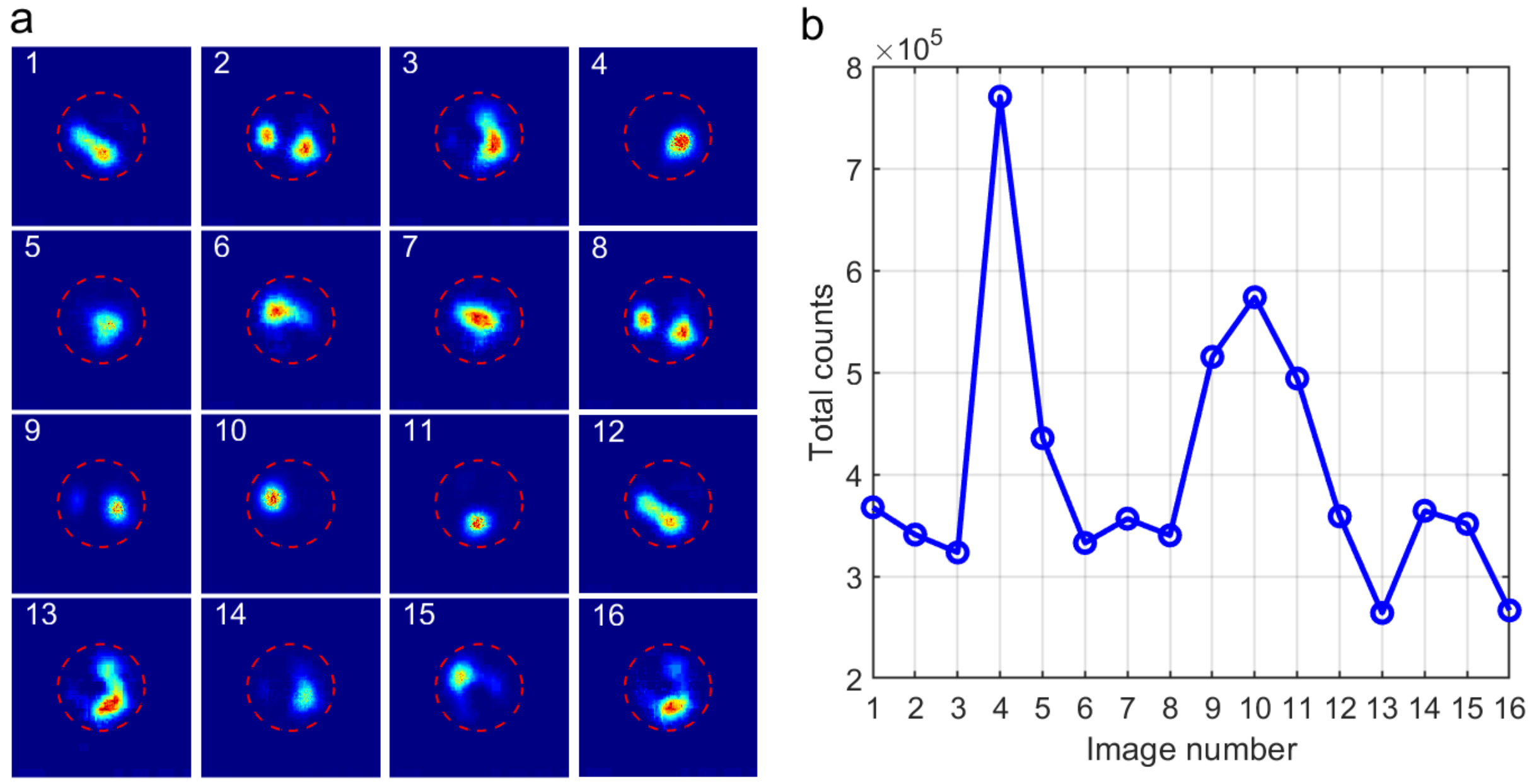


Figure S5: **Representative frames sampled during a fixed-$V_x$ voltage-scan window.** Sixteen representative frames extracted from the selected time interval are shown (a) at fixed $V_x$ together with the corresponding total-count trace (b). The dashed circle marks the multimode-core region used for energy integration. The sequence illustrates that coarse voltage scans sample a range of localized and intermediate interference states, including abrupt changes in spatial localization and total intensity.

measurement.

This temporal redundancy provides a direct measure of stability. For a fixed voltage configuration, stable basins manifest as plateau regions in energy redistribution traces, with minimal frame-to-frame variation. In contrast, unstable or transient configurations exhibit fluctuations or systematic drift, reflecting sensitivity to perturbations. Occasional abrupt jumps in the localized output during the scan correspond to crossings between neighbouring basins, providing additional signatures of basin boundaries and sensitivity to specific control directions. This intrinsic stability filtering enables identification of robust operating basins directly from the measured data.

The sampled phase space reveals a structured landscape of accessible states. Localized output modes (Spots 1–3) emerge as extended regions in control space, corresponding to basins of attraction. These basins exhibit varying efficiency, spatial isolation, and temporal stability, reflecting the influence of system imperfections such as fabrication tolerances, modal imbalance, and residual coupling.

Among the observed states, Spot 1 exhibits both the highest efficiency and strongest temporal

stability, consistent with its identification as the most robust basin in the main text. Partial overlap between states indicates proximity of basins and possible transition pathways, while deviations from ideal symmetry reflect non-idealities in the photonic lantern.

This protocol demonstrates that coarse phase-space sampling combined with temporal analysis provides a direct and experimentally accessible method for identifying basin structure, robustness, and operating regimes in multimode systems under phase-only control.


\end{document}